\documentclass[journal]{IEEEtran}
\usepackage{cite}

\ifCLASSINFOpdf
  \usepackage[pdftex]{graphicx}
\else
\fi
\usepackage{amsmath}
\ifCLASSOPTIONcompsoc
  \usepackage[caption=false,font=normalsize,labelfont=sf,textfont=sf]{subfig}
\else
  \usepackage[caption=false,font=footnotesize]{subfig}
\fi
\usepackage{amssymb}
\usepackage[nolist]{acronym}
\usepackage{bm}
\usepackage{mathtools}

\DeclarePairedDelimiter\floor{\lfloor}{\rfloor}
\usepackage{units}
\usepackage[noabbrev,capitalise]{cleveref}
\crefname{appsec}{Appendix}{Appendices}
\crefname{equation}{}{}
\usepackage{booktabs}
\usepackage{color}

\begin{document}
\begin{acronym}
	\acro{PDF}{probability density function}
\acro{DoA}{direction-of-arrival}
\acrodefplural{DoA}[DoAs]{directions-of-arrival}
\acro{RSS}{received signal strength}
\acro{RHCP}{right hand circular polarization}
\acro{WSN}{wireless sensor networks}
\acro{MUSIC}{multiple signal characterization}
\acro{ML}{maximum likelihood}
\acro{FIM}{Fisher information matrix}
\acro{CRB}{Cram\'{e}r-Rao bound}
\acro{ULA}{uniform linear array}
\acro{UCA}{uniform circular array}
\acro{EMF}{electromagnetic field}
\acro{MEMS}{microelectromechanical system}
\acro{GNSS}{global navigation satellite system}
\acro{WSN}{wireless sensor networks}
\acro{SNR}{signal-to-noise ratio}
\acro{FOV}{field of view}
\acro{MMA}{multi-mode antenna}
\acro{EADF}{effective aperture distribution function}
\acro{TCM}{theory of characteristic modes}
\acro{MIMO}{multiple-input multiple-output}
\acro{5G}{fifth generation}
\acro{UAV}{unmanned aerial vehicle}
\acro{TDMA}{time-division multiple access}
\acro{RMSE}{root-mean-square error}
\acro{RC}{reduced complexity}
\acro{MIMO}{multiple-input, multiple-output}
\acro{ESPRIT}{estimation of signal parameters by rotational invariance techniques}
\acro{IQML}{iterative quadratic maximum likelihood}
\acro{AIT}{array interpolation technique}
\acro{WM}{wavefield modeling}
\acro{URA}{uniform rectangular array}
\acro{ToF}{time-of-flight}
\acro{FoV}{field of view}
\end{acronym}
%
\title{On the Potential of Multi-Mode Antennas for Direction-of-Arrival Estimation}
%
%
%

\author{Robert~P\"ohlmann,~
        Sami~Alkubti~Almasri,~
        Siwei~Zhang,~
        Thomas~Jost,~
        Armin~Dammann~
        and~Peter A. Hoeher
\thanks{Robert P\"ohlmann, Siwei Zhang, Thomas Jost and Armin Damman are with the Institute of Communications and Navigation, German Aerospace Center (DLR), Wessling 82234, Germany. (e-mail: {Robert.Poehlmann, Siwei.Zhang, Thomas.Jost, Armin.Dammann}@dlr.de)}
\thanks{Sami Alkubti Almasri and Peter A. Hoeher are with University of Kiel, Faculty of Engineering, Kiel 24143, Germany. (e-mail: {saaa, ph}@tf.uni-kiel.de)}
}

\maketitle

\begin{abstract}
A \ac{MMA} can be an interesting alternative to a conventional phased antenna array for \ac{DoA} estimation. By \ac{MMA} we mean a single physical radiator with multiple ports, which excite different characteristic modes. In contrast to phased arrays, a closed-form mathematical model of the antenna response, like a steering vector, is not straightforward to define for \acp{MMA}. Instead one has to rely on calibration measurement or \ac{EMF} simulation data, which is discrete. To perform \ac{DoA} estimation, an \ac{AIT} and \ac{WM} are suggested as methods with inherent interpolation capabilities, fully taking antenna nonidealities like mutual coupling into account. We present a non-coherent \ac{DoA} estimator for low-cost receivers and show how coherent \ac{DoA} estimation and joint \ac{DoA} and polarization estimation can be performed with \acp{MMA}. Utilizing these methods, we assess the \ac{DoA} estimation performance of an \ac{MMA} prototype in simulations for both 2D and 3D cases. The results show that \ac{WM} outperforms \ac{AIT} for high SNR. The coherent estimation is superior to non-coherent, especially in 3D, because non-coherent suffers from estimation ambiguities.
In conclusion, \ac{DoA} estimation with a single \ac{MMA} is feasible and accurate.
\end{abstract}


%
\IEEEpeerreviewmaketitle

\acresetall
\section{Introduction}\label{s:introduction}
%
%
%
%
%
%
\IEEEPARstart{M}{ulti}-mode antennas leverage the \ac{TCM}, which was originally published in the 1970s \cite{garbacz1971,harrington1971a}. However it tended to be sidelined until the 2000s, when the need to fit antennas into compact handheld devices arose. It has then been realized that for efficient radiation, the dominating characteristic mode of the device chassis has to be excited \cite{villanen2006}. The \ac{TCM} has since then received an increasing amount of attention within the antenna community \cite{lau2016}, as it allows assessment of the radiation characteristics of defined shapes in terms of individual characteristic modes. An introduction to the concept of characteristic modes can be found in \cite{chen2015,cabedo-fabres2007}. With \ac{MIMO} communication systems becoming popular, the next step in the application of the \ac{TCM} was to excite multiple characteristic modes. Our definition of a \ac{MMA} is thus as a multiport antenna, where different characteristic modes are excited independently, see e.g. \cite{martens2011a,zhao2018}. Using \acp{MMA} allows to develop compact antennas for \ac{MIMO} systems. Finally, multiple \acp{MMA} can be grouped into an array, forming a multi-mode, multi-element antenna which can serve at a base station for ultra-high data rates \cite{manteuffel2016}. So far, design and application of \acp{MMA} was focused on communication applications only.

Contrarily, this paper highlights the usage of \acp{MMA} for positioning purposes, specifically for \ac{DoA} estimation \cite{pohlmann2017,pohlmann2017b,almasri2017}. \ac{DoA} is an essential part for numerous applications like robust \ac{GNSS} receivers \cite{heckler2011}, multipath assisted positioning \cite{gentner2016a} and channel modeling \cite{jost2012}. While \acp{MMA} could be used for all of these applications, we highlight two potential applications for \ac{DoA} where \acp{MMA} are especially suited. First, \ac{5G} mobile networks are expected to provide high-throughput together with location information as a service \cite{peral-rosado2017,koivisto2017}. \ac{5G} is also envisaged to leverage location information to improve communication \cite{taranto2014}. A wideband multi-mode, multi-element antenna, like the one from \cite{manteuffel2016}, could be applied at the basestation side. A second application where \acp{MMA} are well suited are multi-agent robotic systems \cite{sahin2008}, which are envisaged e.g. for terrestrial surveillance, disaster management and extra-terrestrial exploration. When it comes to small \acp{UAV} like quadrocopters, stringent size, weight and shape constraints apply, making the design of multi antenna systems challenging. Here the \ac{TCM} offers a handy tool to use the \ac{UAV} structure for radiation, see e.g. \cite{chen2014,sow2017}. With a single port antenna, it is only possible to obtain range information by measuring the signal \ac{ToF}. It has been shown that angular information, in addition to range information, is very valuable for autonomous navigation of multi-agent robotic systems, as it enables orientation estimation and makes positioning more robust \cite{pohlmann2018}. Going one step further and applying the \ac{TCM} to construct a multiport antenna, what we call \ac{MMA}, would allow \acp{DoA} estimation and thus make angular information available.

The design of antenna arrays for \ac{DoA} estimation and beamforming is well known \cite{balanis2007a}, while \acp{MMA} have not been widely considered for this purpose yet. The antenna response of an \ac{MMA} cannot be simply described by a steering vector. The plethora of methods known from array signal processing and \ac{DoA} estimation \cite{krim1996,tuncer2009} can thus not be directly applied to \acp{MMA}. Instead, one has to rely on either \ac{WM} and manifold separation or the \ac{AIT} to model the \ac{MMA} response.
Non-coherent \cite{pohlmann2017} and coherent \cite{pohlmann2017b} \ac{DoA} estimation for \acp{MMA} by \ac{WM}, and coherent \ac{DoA} estimation by \ac{AIT} \cite{almasri2017} have been introduced. In this paper we compare the \ac{AIT} and \ac{WM} approaches and extend the \ac{DoA} estimation scheme to include polarization. Moreover we introduce a new non-coherent \ac{DoA} estimator with reduced complexity and briefly analyze the differences between non-coherent and coherent \ac{DoA} estimation in terms of ambiguities for a specific \ac{MMA} design.

Instead of exploiting the \ac{TCM} to excite different modes on a single radiator, another design approach is to collocate multiple antennas within a small footprint \cite{konanur2005,chiu2007,elnour2010}. Similar to an \ac{MMA}, one obtains a multiport antenna with different radiation patterns for each port. With collocated antennas it is also possible to achieve low mutual coupling between the ports \cite{sarrazin2010}. While the design approach of the antenna is different, its pattern can also be described by the generic models introduced in this paper, utilizing \ac{AIT} or \ac{WM}. Therefore the described methods for \ac{DoA} estimation can be applied not only to \acp{MMA}, but also to collocated antennas.

\begin{figure}[t]
	\centering
	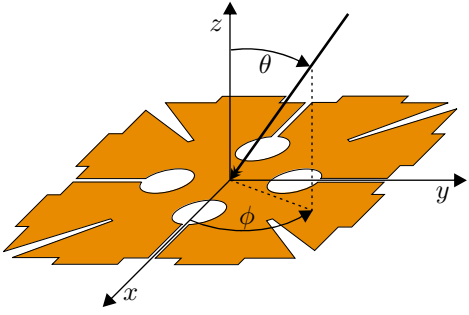
	\caption{Single \ac{MMA} \cite{manteuffel2016} in x-y-plane, coordinate system and incoming signal.}
	\label{fig:coordinates_mma}
\end{figure}
\begin{figure}[t]
	\centering
	\includegraphics{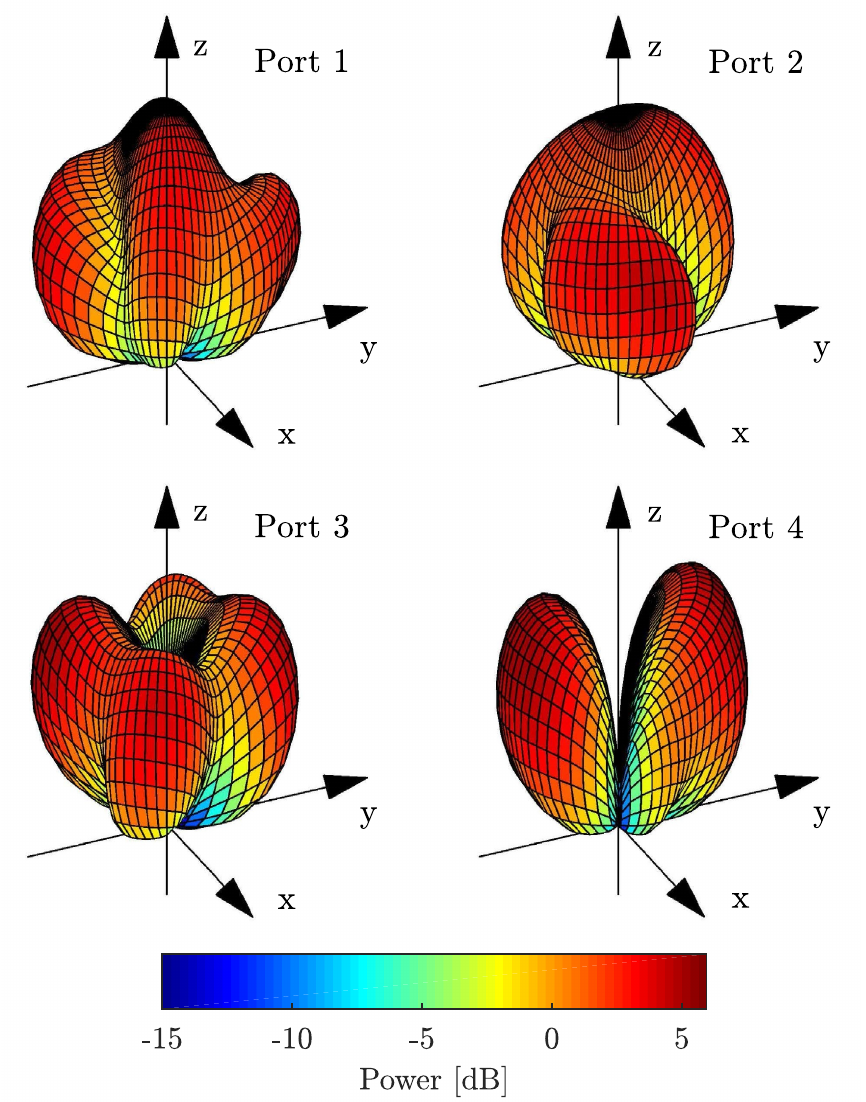}
	\caption{3D power patterns for \acf{RHCP} of the investigated \ac{MMA} \cite{manteuffel2016}.}
	\label{fig:Power_3D_RHCP}
\end{figure}

The aim of this paper is threefold. First we want to highlight the potential of \acp{MMA} for \ac{DoA} estimation. Second we present suitable methods for determining the \ac{DoA} with \acp{MMA}. And third we analyze the \ac{DoA} estimation performance using \ac{EMF} simulation data of an \ac{MMA} prototype. To this end we present two fundamentally different approaches how an \ac{MMA} can be modeled for signal processing, see \cref{s:mmamodel}. Both approaches take antenna nonidealities like mutual coupling into account. In \cref{s:sp} we introduce a non-coherent, i.e. \ac{RSS} measurement based \ac{DoA} estimation scheme, aiming at low-cost and low-complexity receivers. In addition to the \ac{ML} estimator, we also develop a low-complexity alternative. We further present a coherent \ac{DoA} estimator, which is the standard approach and suitable for e.g. navigation of multi-agent robotic systems or roust \ac{GNSS} receivers. Finally we extend the \ac{DoA} estimation approach to jointly estimate the polarization, increasing robustness in case the polarization is unknown. Joint \ac{DoA} and polarization estimation is also useful for applications like channel modeling. The analysis is based on \ac{EMF} simulation data from an \ac{MMA} prototype presented in \cite{manteuffel2016}. While \cite{manteuffel2016} shows an array of \acp{MMA}, this paper focuses on a single \ac{MMA}. The dimension of the antenna is $\unit[30]{mm} \times \unit[30]{mm}$, which is $0.725 \lambda \times 0.725 \lambda$ at the center frequency \unit[7.25]{GHz}. A drawing of the \ac{MMA} can be seen in \cref{fig:coordinates_mma} and the 3D power patterns of the four ports are shown in \cref{fig:Power_3D_RHCP}. The envelope correlation between all ports is below $-\unit[20]{dB}$. For more details about the antenna, please refer to \cite{manteuffel2016}. In \cref{s:performance} we show the \ac{DoA} estimation performance for 2D and 3D respectively. Finally in \cref{s:discussion} we discuss pros and cons of the two presented antenna response models based on \ac{AIT} and \ac{WM}. We also talk about the choice of basis functions for the \ac{WM} approach. Finally we give hints about the practical implementation of the proposed methods.

Throughout the paper, we use the following notation:
\begin{itemize}
	\item Vectors are written in bold lowercase letters and matrices in bold capital letters.
	\item $(\cdot)^T$, $(\cdot)^H$ stands for vector or matrix transpose and conjugate transpose.
	\item $[\bm{A}]_{i,j}$ refers to the element in row $i$ and column $j$.
	\item $||\bm{A}||$ is the Frobenius norm of matrix $\bm{A}$.
	\item $\bm{A} \odot \bm{B}$ is the Hadamard-Schur product of matrices $\bm{A}$ and $\bm{B}$.
	\item $\bm{A} \otimes \bm{B}$ is the Kronecker product of matrices $\bm{A}$ and $\bm{B}$.
	\item $\bm{A}^\dagger$ is the Moore-Penrose pseudoinverse of matrix $\bm{A}$.
	\item $\operatorname{tr}\{\bm{A}\}$ and $\det\{\bm{A}\}$ are the trace and determinant of matrix $\bm{A}$.
	\item $\mathbb{I}_N$ is an $N \times N$ identity matrix.
	\item $\bm{1}_N = [1, ..., 1]^T$ is a vector of ones with length $N$.
	\item $\operatorname{E}\{\cdot\}$, $\operatorname{var}\{\cdot\}$ denote expectation and variance.
	\item $\operatorname{cov}\{\cdot,\cdot\}$ is the covariance matrix.
	\item $\operatorname{Re}\{\cdot\}$, $\operatorname{Im}\{.\}$ refer to real and imaginary part.
	\item $\floor{x}$ is the floor function returning the greatest integer less than or equal to $x$.
\end{itemize}

\section{Multi-Mode Antenna Response Vector}\label{s:mmamodel}
To perform \ac{DoA} estimation, we require a continuous, closed-form expression for the antenna response
\begin{equation}
	a_m(\theta,\phi) = \sqrt{g_m(\theta,\phi)} e^{\mathrm{j} \Phi_m(\theta,\phi)},
\end{equation}
for antenna port $m=1,...,M$, antenna gain $g_m(\theta,\phi)$ and antenna phase response $\Phi_m(\theta,\phi)$  \cite{balanis2007a}. Inclination angle $\theta$ and azimuth angle $\phi$ are visualized in \cref{fig:coordinates_mma}. The antenna response vector for $M$ ports is defined as
\begin{equation}\label{eq:arv}
	\bm{a}(\theta,\phi) = 
	\begin{bmatrix}
		a_1(\theta,\phi) & ... & a_{M}(\theta,\phi)
	\end{bmatrix}^T.
\end{equation}
Please note that $\bm{a}(\theta,\phi)$ represents in general a non-linear vector function taking all effects into account. For \acp{MMA}, in contrast to ideal antenna arrays consisting of isotropic antennas, it is not straightforward to find an analytical expression. Instead, the starting point for determining $\bm{a}(\theta,\phi)$ for an \ac{MMA} are spatial samples of the antenna response given by
\begin{equation}
	\bm{e}_q = \begin{bmatrix}
		e_{q,1} & ... & e_{q,M}
	\end{bmatrix}^T
\end{equation}
for a specific sampling point $\{\theta_q,\phi_q\}$. For the entire sphere this extends to $\bm{E} = [\bm{e}_1, ..., \bm{e}_{Q}]$ with $Q$ total sampling points. The spatial samples obtained by antenna calibration measurement or \ac{EMF} simulation are inherently discrete, hence an interpolation strategy is needed. The goal is to find a closed-form expression for $\bm{a}(\theta,\phi)$ such that
\begin{equation}
	\bm{a}(\theta_q,\phi_q) \approx \bm{e}_q, \quad \forall \: q \in \lbrace 1, ..., Q \rbrace
\end{equation}
holds and $\bm{a}(\theta,\phi)$ is continuous in $\theta$ and $\phi$. 

\subsection{Array Interpolation Technique}\label{ss:ait}
The idea of \ac{AIT} is to model $\bm{a}(\theta,\phi)$ as
\begin{equation}
	\bm{a}(\theta,\phi) = \bm{H} \, \bm{a}_{\mathrm{ideal}}(\theta,\phi)
\end{equation}
being a linear transformation of the response of a virtual, ideal array\footnote{The ideal array response is obtained assuming isotropic antennas without mutual coupling such that each element of $\bm{a}_{\mathrm{ideal}}$ has unit magnitude and its phase directly depends on the geometrical relation between the incoming wave and the antenna position within the array aperture. Therefore knowing the antenna positions, $\bm{a}_{\mathrm{ideal}}$ can be calculated in a straightforward manner.}
$\bm{a}_{\mathrm{ideal}}(\theta,\phi)$. The linear transformation is described by the interpolation matrix $\bm{H} \in \mathbb{C}^{M \times M}$. \ac{AIT} was first proposed in \cite{bronez1988} and has been extended in e.g. \cite{friedlander1990,friedlander1992}. Defining $\bm{A}_{\mathrm{ideal}} = [\bm{a}_{\mathrm{ideal}}(\theta_1,\phi_1), ..., \bm{a}_{\mathrm{ideal}}(\theta_{Q},\phi_{Q})]$, the mapping can be found by solving the optimization problem
\begin{equation}\label{eq:aitoptfull}
	 \lbrace \hat{\bm{H}}, \hat{\bm{a}}_{\mathrm{ideal}}(\theta,\phi) \rbrace = \arg\underset{\bm{H}, \bm{a}_{\mathrm{ideal}}(\theta,\phi)}{\min} \left|\left| \bm{H}^H \bm{A}_{\mathrm{ideal}} - \bm{E} \right|\right|^2.
\end{equation}
Optimizing $\bm{a}_{\mathrm{ideal}}$ means moving the elements of the virtual array to an optimum position.
Although this can be done, their position is often chosen heuristically \cite{buhren2004}. By that the optimization problem in \cref{eq:aitoptfull} simplifies to an optimization of $\bm{H}$. From an algorithmic point of view, it is advantageous to work with uniform arrays. In order to apply \ac{AIT} for \acp{MMA} in the 2D case, we assume a cut through the x-z-plane and apply a \ac{ULA} oriented along the x-axis
\begin{equation}\label{eq:ula}
	a_{\mathrm{ULA},m}(\theta) = e^{\mathrm{j} \frac{2 \pi}{\lambda_{\mathrm{c}}} (m-1) d \sin(\theta)}
\end{equation}
with inter-element spacing $d=\lambda_{\mathrm{c}}/4$ for wavelength $\lambda_{\mathrm{c}}$, yielding the steering vector
\begin{equation}
	\bm{a}_{\mathrm{ULA}}(\theta) =
	\begin{bmatrix}
		a_{\mathrm{ULA},1}(\theta) & ... & a_{\mathrm{ULA},{M}}(\theta)
	\end{bmatrix}^T.
\end{equation}
For 3D, we limit ourselves to a \ac{URA}. Assuming the \ac{URA} lies on the x-y-plane, we have
\begin{subequations}
\begin{equation}
	a_{\mathrm{x},m}(\theta,\phi) = e^{\mathrm{j} \frac{2\pi}{\lambda_{\mathrm{c}}} (m-1) d \sin(\theta)\cos(\phi)},
\end{equation}
\begin{equation}
	a_{\mathrm{y},m}(\theta,\phi) = e^{\mathrm{j} \frac{2\pi}{\lambda_{\mathrm{c}}} (m-1) d \sin(\theta)\sin(\phi)}.
\end{equation}
\end{subequations}
Defining
$\bm{a}_x(\theta,\phi) = [a_{\mathrm{x},1}(\theta,\phi), ..., a_{\mathrm{x},M}(\theta,\phi)]^T$ and $\bm{a}_y(\theta,\phi) = [a_{\mathrm{y},1}(\theta,\phi), ..., a_{\mathrm{y},M}(\theta,\phi)]^T$, we obtain the steering vector of the \ac{URA}
\begin{equation}\label{eq:ura}
	\bm{a}_{\mathrm{URA}}(\theta,\phi) = \bm{a}_{\mathrm{x}}(\theta,\phi) \otimes \bm{a}_{\mathrm{y}}(\theta,\phi).
\end{equation}
As these ideal responses and a real \ac{MMA} response are in general quite different, usually no linear mapping can be found which represents the original antenna response with sufficient accuracy. Instead the manifold is divided into equally sized sectors, such that for each sector $(c)$ we have samples of the antenna response $\left\{\bm{e}_1^{(c)},...,\bm{e}_{Q^{(c)}}^{(c)}\right\} \subset \left\{\bm{e}_1,...,\bm{e}_{Q}\right\}$, sampled at points
$\left\{ \big\{\theta_1^{(c)},\phi_1^{(c)}\big\}, ...,\big\{\theta_{Q^{(c)}}^{(c)},\phi_{Q^{(c)}}^{(c)}\big\}\right\} \subset \left\{\big\{\theta_1,\phi_1,\big\},...,\big\{\theta_Q,\phi_Q\big\}\right\}$, where samples and sampling points of the respective sector are subsets of all available samples and sampling points.
The sector-wise minimization problem, using the ideal antenna response
$\bm{A}_{\mathrm{ideal}}^{(c)} = \left[\bm{a}_{\mathrm{ideal}}\left(\theta_1^{(c)},\phi_1^{(c)}\right), ..., \bm{a}_{\mathrm{ideal}}\left(\theta_{Q^{(c)}}^{(c)},\phi_{Q^{(c)}}^{(c)}\right)\right]$ and the samples of the antenna response $\bm{E}^{(c)} = [\bm{e}_1^{(c)}, ..., \bm{e}_{Q}^{(c)}]$, then becomes
\begin{equation}
	\hat{\bm{H}}^{(c)} = \arg\underset{\bm{H}^{(c)}}{\min} \left|\left| \left(\bm{H}^{(c)}\right)^H \bm{A}_{\mathrm{ideal}}^{(c)} - \bm{E}^{(c)} \right|\right|^2,
\end{equation}
which can be solved by a least squares approach
\begin{equation}
	\bm{H}^{(c)} = \bm{E}^{(c)} \left(\bm{A}_{\mathrm{ideal}}^{(c)}\right)^H \left( \bm{A}_{\mathrm{ideal}}^{(c)} \left(\bm{A}_{\mathrm{ideal}}^{(c)}\right)^H \right)^{-1}.
\end{equation}
We only consider linear \ac{AIT} here, as it is simple and has low complexity. Nonlinear techniques with higher computational cost exist as well \cite{marinho2018}.

\subsection{Wavefield Modeling and Manifold Separation}\label{ss:wms}
Another possibility to perform the interpolation is by building on a technique called wavefield modeling and manifold separation \cite{doron1994,costa2010}. The key finding here is that the antenna response vector is modeled as
\begin{equation}\label{eq:ms}
	\bm{a}(\theta,\phi) = \bm{G} \, \bm{b}(\theta,\phi) \in \mathbb{C}^{M}
\end{equation}
and can be decomposed into a product of the sampling matrix $\bm{G} \in \mathbb{C}^{M \times U}$, which is independent of the wavefield, i.e. the \ac{DoA}, and the basis vector $\bm{b}(\theta,\phi) \in \mathbb{C}^{U}$, which is independent of the antenna \cite{doron1994}. This decomposition requires the $U$ basis functions to be orthonormal on the antenna manifold $\theta \in [-\pi,\pi)$ for 2D or $\theta \in [0,\pi], \: \phi \in [0,2\pi)$ for 3D respectively. For 2D we assume a cut through the x-z-plane. The antenna response vector $\bm{a}(\theta,\phi)$ must also be square integrable on the manifold. A suitable basis for 2D is given by the Fourier functions
\begin{equation}\label{eq:fourier}
	\bm{b}(\theta) = \frac{1}{\sqrt{2\pi}} e^{\mathrm{j}\theta u_\theta}, \: u_\theta = 		\floor*{-\frac{U-1}{2}}, ..., 0, ..., \floor*{\frac{U-1}{2}}.
\end{equation}
For 3D the spherical harmonic functions
\begin{equation}\label{eq:Ylm}
	Y_l^m(\theta,\phi) = \sqrt{\frac{2l+1}{4\pi}\frac{(l-m)!}{(l+m)!}} P_l^m(\cos(\theta)) e^{\mathrm{j}m\phi},
\end{equation}
with degree $l \in \lbrace 0,...,L \rbrace$ for maximum degree $L$ and order $m \in \lbrace -l,...,l \rbrace$ fulfill the orthonormality property \cite{olver2010}. Please note that we use $l$ and $m$ here to be consistent with the literature, $m$ is not to be confused with the antenna index utilized in the rest of this paper. $P_l^m(\cdot)$ is the associated Legendre polynomial, see \cref{eq:Plm} in \cref{a:sh}. Defining $Y_u(\theta,\phi)$ analogous to $Y_l^m(\theta,\phi)$ with the enumeration $u=(l+1)l+m+1$ for $u=1,...,U$, we can form a basis
\begin{equation}\label{eq:sh}
	\bm{b}(\theta,\phi) =
	\begin{bmatrix}
		Y_1(\theta,\phi) & ... & Y_{U}(\theta,\phi)
	\end{bmatrix}^T.
\end{equation}
Another choice would be the 2D Fourier functions
\begin{subequations}
\begin{equation}
	\resizebox{.9\hsize}{!}{
		$\bm{b}(\theta) = \frac{1}{\sqrt{2\pi}} e^{\mathrm{j}\theta u_\theta} \: u_\theta = \floor*{-\frac{\sqrt{U}-1}{2}}, ..., 0, ..., \floor*{\frac{\sqrt{U}-1}{2}},$
	}
\end{equation}
\begin{equation}
	\resizebox{.9\hsize}{!}{
		$\bm{b}(\phi) = \frac{1}{\sqrt{2\pi}} e^{\mathrm{j}\phi u_\phi}, \: u_\phi = 		\floor*{-\frac{\sqrt{U}-1}{2}}, ..., 0, ..., \floor*{\frac{\sqrt{U}-1}{2}},$
	}
\end{equation}
\end{subequations}

leading to the basis vector
\begin{equation}\label{eq:eadf}
	\bm{b}(\theta,\phi) = \bm{b}_\theta(\theta) \otimes \bm{b}_\phi(\phi).
\end{equation}
We chose this definition, essentially limiting $U$ to square numbers, to unify the definition of $U$ for \cref{eq:fourier}, \cref{eq:sh} and \cref{eq:eadf}. In practice, this limitation is not necessary and the number of coefficients in $\theta$ and $\phi$ domain may also be different. The approach \cref{eq:eadf} is likewise called \ac{EADF} \cite{belloni2007}. The 2D Fourier functions are orthonormal on the torus, not on the sphere, i.e. the data has to be expanded to be periodic in both inclination and azimuth \cite{costa2010}. We provide a short discussion on the choice of basis functions in \cref{s:discussion}.

In \cite{doron1994} it is shown that when the number of coefficients $U$ is increased, the magnitude of the entries in $G$ decays superexponentially for $|u_\theta| > \kappa R_{\mathrm{s}}$, $|u_\phi| > \kappa R_{\mathrm{s}}$ and  $l > \kappa R_{\mathrm{s}}$, where $\kappa=\frac{2\pi}{\lambda_{\mathrm{c}}}$ is the wavenumber and $R_{\mathrm{s}}$ is the radius of the smallest sphere enclosing the antenna. From this observation a rule of thumb can be deduced, that the expansion can be truncated at $2 \kappa R_{\mathrm{s}}$,
while an accurate representation of the antenna response can be preserved \cite{doron1994}. Following this rule of thumb leads to $U  \approx 4 \kappa R_{\mathrm{s}} + 1$ coefficients for the Fourier series, $U \approx 8 \kappa^2 R_{\mathrm{s}}^2 + 4 \kappa R_{\mathrm{s}} +1$ for spherical harmonics and $U \approx (4 \kappa R_{\mathrm{s}} + 1)^2$ for the 2D Fourier series. In practice, $U$ can also be adjusted according to the noise floor of the calibration measurements. In that case, $U$ is chosen such that the sampling matrix is truncated one magnitude above the noise floor \cite{belloni2007}.

Using $\bm{E}$, we can determine the sampling matrix $\bm{G}$ for a given basis $\bm{B} = \left[\bm{b}(\theta_1,\phi_1), ..., \bm{b}(\theta_Q,\phi_Q)\right]$ by least squares as
\begin{equation}\label{eq:wmls}
	\hat{\bm{G}} =  \bm{E} \bm{B}^H \left( \bm{B} \bm{B}^H \right)^{-1}.
\end{equation}
For this equation to be solvable, in general $M \leq U \leq Q$. When a regular grid is employed, which is nonuniform on the sphere but often used for antenna measurements, it should be ensured that $U \ll Q$ \cite{hansen1988}. Once $\hat{\bm{G}}$ has been found, the interpolation can be performed by \cref{eq:ms}. For basis functions \cref{eq:fourier,eq:eadf}, \cref{eq:wmls} is equivalent to performing a Fourier transform.

\subsection{Antenna Characteristics}\label{ss:ac}
\begin{figure}[t]
	\centering
	\includegraphics{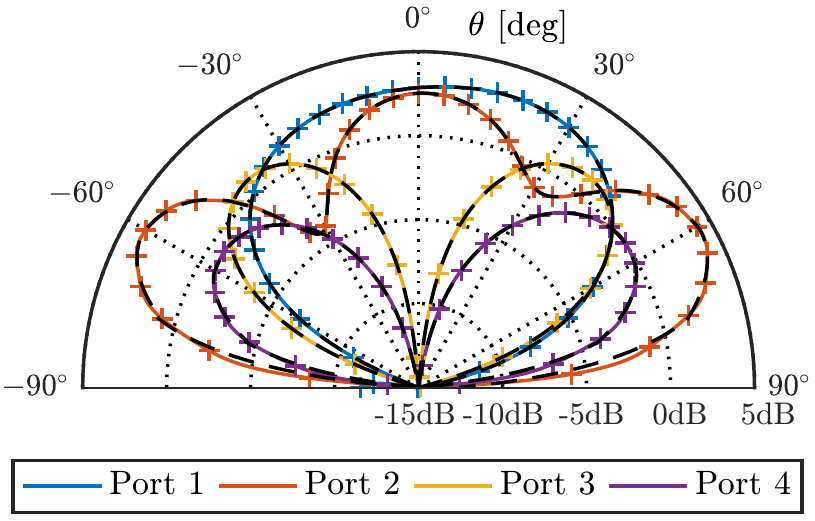}
	\caption{Sampled (crosses) and interpolated with \ac{WM} (solid lines) and \ac{AIT} (black dashed lines) x-z-plane \ac{MMA} power pattern for \ac{RHCP}.}
	\label{fig:Power_2D_RHCP}
\end{figure}

\begin{figure}[t]
	\centering
	\includegraphics{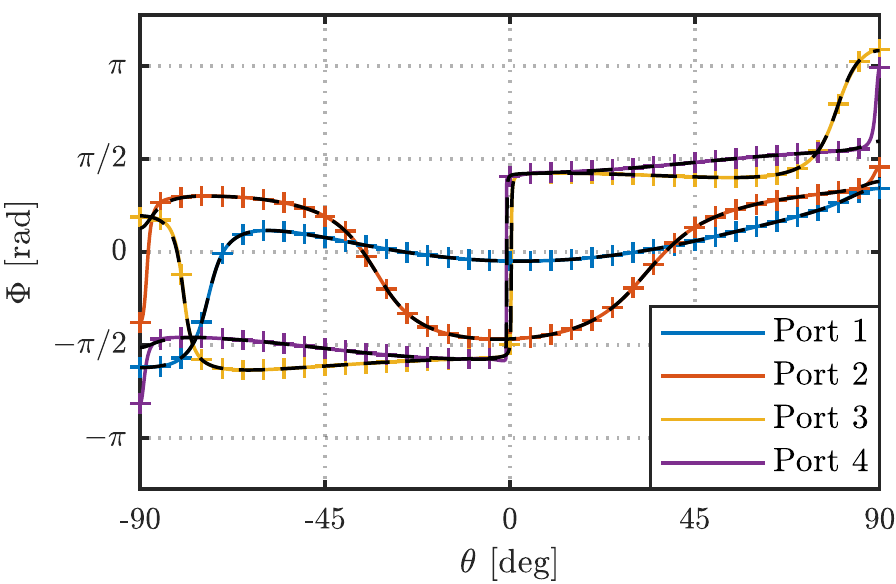}
	\caption{Sampled (crosses) and interpolated with \ac{WM} (solid lines) and \ac{AIT} (black dashed lines) x-z-plane \ac{MMA} phase pattern for \ac{RHCP}.}
	\label{fig:Phase_2D_RHCP}
\end{figure}

To illustrate the interpolation process, we show 2D cuts of both power, \cref{fig:Power_2D_RHCP}, and phase patterns, \cref{fig:Phase_2D_RHCP}, of the \ac{MMA} prototype for a fixed polarization. The figures present both discrete \ac{EMF} simulation data and interpolated patterns with \ac{AIT} and \ac{WM}. For \ac{AIT} we assume a virtual \ac{ULA} with $\lambda_{\mathrm{c}}/4$ spacing and four elements. The manifold is divided into $30^\circ$ sectors with $15^\circ$ overlap, yielding 11 matrices $\hat{\bm{H}}^{(c)} \in \mathbb{C}^{4\times4}$, i.e. $44$ weighting factors per antenna port. As basis for \ac{WM} we use Fourier functions \cref{eq:fourier} with  $U = 15 \approx 4 \kappa R_{\mathrm{s}} + 1$ coefficients per port. The basis for our analysis is noise-free data obtained by \ac{EMF} simulation, therefore we choose a large $U$ to achieve exact interpolation. When measurement data from an anechoic chamber is used, $U$ can be significantly reduced, as measurement data is always noisy. It can be seen from \cref{fig:Power_2D_RHCP,fig:Phase_2D_RHCP} that both \ac{AIT} and \ac{WM} accurately interpolate in power and phase domain. For \ac{AIT}, a slight deviation for low elevations, i.e. $|\theta| > 70^\circ$, is visible. In \cref{ss:2d} both approaches are compared in terms of their impact on \ac{DoA} estimation. \Cref{fig:Power_3D_RHCP} shows the interpolated power pattern of the \ac{MMA} in 3D. Obviously the different ports of the \ac{MMA} have distinct characteristics. These are utilized by the signal processing schemes presented in the next section to estimate the \ac{DoA} of incoming signals.

\section{DoA Estimation with Multi-Mode Antennas}\label{s:sp}
In this section we derive estimators for non-coherent and coherent \ac{DoA} estimation as well as joint \ac{DoA} and polarization estimation. Non-coherent means that the receiver has only knowledge of \ac{RSS} measurements, i.e. power of the received signal. We also provide a fundamental limit in terms of the \ac{CRB} for every estimator. All equations are given for 3D, i.e. azimuth and inclination. Simplification to 2D is straightforward. Using the non-coherent approach, only a single signal can be estimated. Coherent estimation allows to distinguish between multiple overlaying signals \cite{wax1989}. How many signal parameters can be uniquely identified depends on the number of antenna ports, the antenna response, and the correlation of the signals. If the signals are uncorrelated and any subset of $P$ antenna response vectors is linearly independent, $P<M$ signals can be identified.

\subsection{Non-Coherent DoA Estimation}\label{ss:noncoh}
The sampled baseband signal $\bm{r}(n) = [r_1(n), ..., r_M(n)]^T$ received at the $M$ ports of the \ac{MMA} is given by
\begin{equation}\label{eq:signal}
	\bm{r}(n) = \bm{a}(\theta,\phi) s(n) + \bm{w}(n),
\end{equation}
where $\bm{a}(\theta,\phi)$ is the antenna response vector and $s(n)$ is the arriving signal, which is assumed to have a small bandwidth compared to the carrier frequency \cite{balanis2007a,viberg2014}. The noise term $\bm{w}(n) \sim \mathcal{CN}(0, \sigma^2 \mathbb{I}_M)$ is i.i.d. white circular symmetric Gaussian noise with covariance matrix $\sigma^2 \mathbb{I}_M$. Assuming stationarity, the \ac{RSS} estimate of port $m$, time-averaged over $N$ samples, can be calculated by
\begin{equation}\label{eq:Prm}
	\check{r}_m = \frac{1}{N} \sum_{n=1}^{N} |r_m(n)|^2.
\end{equation}
For the first receiver type under consideration, being non-coherent, we assume that only \ac{RSS} measurements $\check{\bm{r}} = [\check{r}_1,...,\check{r}_M]^T$ instead of the actual received signals $\bm{r}(n)$ are available.
In \cref{a:sm} we show that for large $N$, the \ac{RSS} measurements $\check{\bm{r}}$ can be well approximated by a Gaussian distribution $\check{\bm{r}} \sim \mathcal{N}(\tilde{\bm{\mu}}, \tilde{\bm{\Sigma}})$ with mean
\begin{equation}\label{eq:gmu}
	\tilde{\bm{\mu}} = \operatorname{E}\{\check{\bm{r}}\}= \bm{g}(\theta,\phi) \check{s} + \bm{1}_M \sigma^2,
\end{equation}
where $\bm{g}(\theta,\phi) = [g_1(\theta,\phi), ..., g_M(\theta,\phi)]^T$ is the antenna gain vector, $\check{s} = \frac{1}{N} \sum_{n=1}^{N} |s(n)|^2$ is the signal power and  $\sigma^2 = \frac{1}{N} \sum_{n=1}^{N} \operatorname{E}\lbrace |w_m(n)|^2 \rbrace$ the noise power. The entries of the diagonal covariance matrix $\tilde{\bm{\Sigma}} = \operatorname{cov}\{\check{\bm{r}},\check{\bm{r}}\}$ are given by
\begin{equation}\label{eq:gsigma}
	[\tilde{\bm{\Sigma}}]_{m,m} = N^{-1}\,\sigma^4 + 2\,N^{-1}\,\sigma^2\,\check{s}\,g_m(\theta,\phi).
\end{equation}
Equations \cref{eq:gmu,eq:gsigma} do not contain the (complex) antenna response, but only the antenna gain. Instead of performing the expansion based on the complex antenna response vector \cref{eq:ms}, we can expand the real antenna gain vector
\begin{equation}\label{eq:realms}
	\bm{g}(\theta,\phi) = \bm{G} \, \bm{b}_{\mathrm{r}}(\theta,\phi) \in \mathbb{R}^{M}.
\end{equation}
The basis $\bm{b}_{\mathrm{r}}(\theta,\phi)$ for Fourier functions is defined analogously to \cref{eq:fourier} or \cref{eq:eadf}, where the negative coefficients are fixed to the complex conjugate of the positive ones. For spherical harmonic functions, the real valued form for defining the basis $\bm{b}_{\mathrm{r}}(\theta,\phi)$ and its derivatives can be found in \cref{a:rsh}. We define the \ac{SNR} with respect to an isotropic antenna with unit gain, i.e.
\begin{equation}
	\text{SNR} = \frac{\check{s}}{\sigma^2}.
\end{equation}
First we consider the general case where both signal power $\check{s}$ and noise power $\sigma^2$ are unknown. The set of parameters to be estimated is then defined by
\begin{equation}\label{eq:noncohun}
	\bm{\zeta} =
	\begin{bmatrix}
		 \theta & \phi & \check{s} & \sigma^2
	\end{bmatrix}^T.
\end{equation}
Neglecting the constant terms, the log-likelihood function is given by
\begin{equation}\label{eq:noncohlh}
	\ln p_{\check{\bm{r}}}(\check{\bm{r}}| \bm{\zeta}) = -\ln\left( \det\{\tilde{\bm{\Sigma}}\} \right) - (\check{\bm{r}}-\tilde{\bm{\mu}})^T \tilde{\bm{\Sigma}}^{-1} (\check{\bm{r}}-\tilde{\bm{\mu}}),
\end{equation}
which leads to the non-coherent \ac{ML} estimator (NC-ML) of
\begin{equation}\label{eq:noncohml}
\begin{split}
	\hat{\bm{\zeta}} &= \arg \max_{\bm{\zeta}} \: \ln p_{\check{\bm{r}}}(\check{\bm{r}}| \bm{\zeta}).
\end{split}
\end{equation}
The variance of any unbiased estimator is lower bounded by the \ac{CRB} \cite{kay1993}. The \ac{CRB} for the non-coherent case (NC-CRB) is given by
\begin{subequations}\label{eq:noncohcrb}
	\begin{equation}
	\operatorname{var}\{\hat{\theta}\} \geq \text{CRB}(\hat{\theta})= [\bm{I}(\bm{\zeta})^{-1}]_{1,1},
	\end{equation}
	\begin{equation}
	\operatorname{var}\{\hat{\phi}\} \geq \text{CRB}(\hat{\phi}) = [\bm{I}(\bm{\zeta})^{-1}]_{2,2},
	\end{equation}
\end{subequations}
with the elements of the Fisher information matrix $\bm{I}(\bm{\zeta}) \in \mathbb{R}^{4 \times 4}$ defined as \cite{kay1993}
\begin{equation}
	[\bm{I}(\bm{\zeta})]_{i,j} = \frac{\partial\tilde{\bm{\mu}}^T}{\partial\zeta_i} \tilde{\bm{\Sigma}}^{-1} \frac{\partial\tilde{\bm{\mu}}}{\partial\zeta_j}
	+ \frac{1}{2} \operatorname{tr}\left\lbrace \tilde{\bm{\Sigma}}^{-1} \frac{\partial\tilde{\bm{\Sigma}}}{\partial\zeta_i} \tilde{\bm{\Sigma}}^{-1} \frac{\partial\tilde{\bm{\Sigma}}}{\partial\zeta_j} \right\rbrace.
\end{equation}

However, solving a non-linear optimization problem with four unknowns is unfavorable for a low-cost and low-complexity receiver. Therefore, we present a \ac{RC} alternative to the \ac{ML} estimator \cref{eq:noncohml}. In practice, the noise power $\sigma^2$ can often be estimated separately, e.g. from unoccupied \ac{TDMA} slots. The unknowns then reduce to
\begin{equation}\label{eq:noncohlcun}
	\bm{\zeta}^\prime =
	\begin{bmatrix}
		\theta & \phi & \check{s}
	\end{bmatrix}^T.
\end{equation}
Neglecting the log-term in \cref{eq:noncohlh} and maximizing the other term we arrive at
\begin{equation}\label{eq:noncohlcnonred}
	\hat{\bm{\zeta}}^\prime = \arg\min_{\bm{\zeta}^\prime} || \check{\bm{r}}^\prime - \bm{g}(\theta,\phi) \check{s}||^2,
\end{equation} 
with $\check{\bm{r}}^\prime = \check{\bm{r}} - \hat{\sigma^2}$. Following the principle from \cite{golub1973a}, we plug in the least-squares estimate $\hat{\check{s}} = \bm{g}(\theta,\phi)^\dagger \check{\bm{r}}^\prime$ and obtain the \ac{RC} estimator (NC-RC)
\begin{equation}\label{eq:noncohlc}
\begin{split}
	\{\hat{\theta},\, \hat{\phi}\} &= \arg\min_{\{\theta,\, \phi\}} || \left(\mathbb{I}_M - \bm{g}(\theta,\phi) \bm{g}(\theta,\phi)^\dagger\right) \check{\bm{r}}^\prime ||^2\\
	&= \arg\min_{\{\theta,\, \phi\}} \operatorname{tr}\left\lbrace \left(\mathbb{I}_M - \bm{g}(\theta,\phi) \bm{g}(\theta,\phi)^\dagger\right) \check{\bm{r}}^\prime \check{\bm{r}}^{\prime T} \right\rbrace,
\end{split}
\end{equation}
where the term in round brackets is idempotent and $^\dagger$ denotes the Moore-Penrose pseudoinverse. The complexity has been reduced from four to two unknowns. The \ac{CRB} for the non-coherent \ac{RC} estimator (NC-RC-CRB) is also given by \cref{eq:noncohcrb}, but with the reduced unknown vector \cref{eq:noncohlcun}. For an efficient implementation, $\left(\mathbb{I}_M - \bm{g}(\theta,\phi) \bm{g}(\theta,\phi)^\dagger\right)$ can be precomputed for a $\theta$-$\phi$ grid with required accuracy.

\subsection{Coherent DoA Estimation}\label{ss:coh}
The signal model for coherent \ac{DoA} estimation, 
\begin{equation}\label{eq:multsignal}
	\bm{r}(n) = \bm{A}(\bm{\theta},\bm{\phi}) \bm{s}(n) + \bm{w}(n),
\end{equation}
is based on \cref{eq:signal}, but is more general since it covers not only one but $P$ signals $\bm{s}(n) =  [s_1(n), ..., s_{P}(n)]^T$, again with small bandwidths compared to the carrier frequency, arriving from different angles $\{\theta_1,\phi_1\},...,\{\theta_P,\phi_P\}$ \cite{balanis2007a,viberg2014}. This leads to $\bm{\theta} =  [\theta_1, ..., \theta_{P}]^T$, $\bm{\phi} =  [\phi_1, ..., \phi_{P}]^T$ and the antenna response vector $\bm{a}(\theta,\phi)$ becomes a matrix,
\begin{equation}
	\bm{A}(\bm{\theta},\bm{\phi}) =
	\begin{bmatrix}
		\bm{a}(\theta_1,\phi_1) & ... & \bm{a}(\theta_{P},\phi_{P})
	\end{bmatrix}.
\end{equation}
It is worth to point out that in the array processing literature, e.g. \cite{krim1996,tuncer2009,viberg2014}, the likelihood functions and estimators for geometric array models like \ac{ULA} or \ac{URA} are well known. In fact, the common model used there is of the same form as \cref{eq:multsignal}. The difference is that $\bm{A}(\bm{\theta},\bm{\phi})$ is called steering matrix and describes phase relationships between the antennas, while in our case $\bm{A}(\bm{\theta},\bm{\phi})$ is the antenna response matrix containing gain and phase information. The equation for the log-likelihood function however remains the same. Ignoring constant terms, it is given by
\begin{equation}\label{eq:cohlh}
\begin{split}
	\ln p_{\bm{r}}(\bm{r}|\bm{\theta}, \bm{\phi}) = &-NM \ln(\sigma^2)\\
	 &- \frac{1}{\sigma^2} \sum_{n=1}^{N} ||\bm{r}(n) - \bm{A}(\bm{\theta},\bm{\phi}) \bm{s}(n) ||^2.
\end{split}
\end{equation}
Based on $N$ received signal samples we can calculate the sample covariance matrix
\begin{equation}
	\hat{\bm{R}}_{\bm{r}} = \frac{1}{N} \sum_{n=1}^{N} \bm{r}(n)  \bm{r}^H(n)
\end{equation}
and obtain the coherent \ac{ML} estimator (C-ML)
\begin{equation}\label{eq:cohml}
	\{\hat{\bm{\theta}},\, \hat{\bm{\phi}}\} = \arg \min_{\{\hat{\bm{\theta}},\, \hat{\bm{\phi}}\}} \operatorname{Re}\{\operatorname{tr}\{\bm{\Pi_{A}}^\perp \hat{\bm{R}_r}\}\},
\end{equation}
with the projector onto the noise subspace $\bm{\Pi}_{\bm{A}}^\perp = \mathbb{I}_M - \bm{A}(\bm{\theta},\bm{\phi})\bm{A}^\dagger(\bm{\theta},\bm{\phi})$. In \cite{delmas2014} it is shown that the \ac{CRB} matrix for the coherent case (C-CRB) can be calculated as
\begin{equation}\label{eq:cohcrb}
	\text{CRB}\left([\hat{\bm{\theta}}^T,\, \hat{\bm{\phi}}^T]^T\right) = \frac{\sigma^2}{2N} \operatorname{Re}\{\bm{D}^H \bm{\Pi_A}^\perp \bm{D} \odot \bm{Z}^T\bm{R}_{\bm{s}}\bm{Z} \}^{-1},
\end{equation}
with
\begin{equation}\label{eq:Dc}
	\bm{D} = 
	\begin{bmatrix}
		\frac{\partial \bm{a}(\theta_1,\phi_1)}{\partial \theta_1} & ... & \frac{\partial \bm{a}(\theta_P,\phi_P)}{\partial \theta_P} & \frac{\partial \bm{a}(\theta_1,\phi_1)}{\partial \phi_1} & ... & \frac{\partial \bm{a}(\theta_P,\phi_P)}{\partial \phi_P}
	\end{bmatrix}
\end{equation}
and the selection matrix
$\bm{Z} = \begin{bmatrix}
		\mathbb{I}_P & \mathbb{I}_P
	\end{bmatrix}$
and $\bm{R}_{\bm{s}} = \frac{1}{N} \sum_{n=1}^{N} \bm{s}(n)\bm{s}^H(n)$.
Depending on which method is used, the derivatives of steering vectors \cref{eq:ula,eq:ura} for \ac{AIT} or Fourier \cref{eq:fourier} and 2D Fourier functions \cref{eq:eadf} for \ac{WM} are trivial. For \ac{WM} with spherical harmonics, derivatives of \cref{eq:Ylm} are provided in \cref{a:sh} for convenience of the reader.

\subsection{Joint DoA and Polarization Estimation}\label{ss:pol}
Different parameterizations describing polarization parameters of electromagnetic waves exist. We use the auxiliary angle $\gamma$ with $0\leq\gamma\leq\frac{\pi}{2}$ and the polarization phase $\beta$ with $-\pi\leq\beta<\pi$ as parameters of the polarization ellipse\footnote{For linearly polarized electromagnetic waves $\beta=0$, for circularly polarized waves $\gamma=\frac{\pi}{4}$ and $\beta=\pm\frac{\pi}{2}$ for left/right hand circular polarization.} \cite{wong2004}. As our \ac{MMA} prototype has different polarizations on different ports, we can apply methods from diversely polarized array processing \cite{swindlehurst1993,costa2012}. We define partial antenna response vectors for a single signal and antenna port $m$
\begin{subequations}
\begin{equation}
	a_{\mathrm{co},m}(\theta,\phi) = \sqrt{g_{\mathrm{co},m}(\theta,\phi)} e^{\mathrm{j} \Phi_{\mathrm{co},m}(\theta,\phi)},
\end{equation}
\begin{equation}
	a_{\mathrm{cross},m}(\theta,\phi) = \sqrt{g_{\mathrm{cross},m}(\theta,\phi)} e^{\mathrm{j} \Phi_{\mathrm{cross},m}(\theta,\phi)}.
\end{equation}
\end{subequations}
where $a_{\mathrm{co},m}(\theta,\phi)$ is the antenna response when being illuminated by a wave with the reference polarization with \ac{DoA} $\{\theta,\phi\}$, while $a_{\mathrm{cross},m}(\theta,\phi)$ results from a wave with orthogonal polarization. 
Correspondingly $g_{\mathrm{co},m}(\theta,\phi)$ and $g_{\mathrm{cross},m}(\theta,\phi)$ are the partial gains and $\Phi_{\mathrm{co},m}(\theta,\phi)$ and $\Phi_{\mathrm{cross},m}(\theta,\phi)$ the partial phase responses.
Forming the partial antenna response vectors
\begin{subequations}
\begin{equation}
	\bm{a}_{\mathrm{co}}(\theta,\phi) =
	\begin{bmatrix}
		a_{\mathrm{co},1}(\theta,\phi) & ... & a_{\mathrm{co},M}(\theta,\phi)
	\end{bmatrix}^T,
\end{equation}
\begin{equation}
	\bm{a}_{\mathrm{cross}}(\theta,\phi) =
	\begin{bmatrix}
		a_{\mathrm{cross},1}(\theta,\phi) & ... & a_{\mathrm{cross},M}(\theta,\phi)
	\end{bmatrix}^T,
\end{equation}
\end{subequations}
the polarimetric antenna response vector is given by
\begin{equation}
	\bm{a}(\theta,\phi,\gamma,\beta) = \sin(\gamma)e^{\mathrm{j}\beta}\bm{a}_{\mathrm{co}}(\theta,\phi) + \cos(\gamma)\bm{a}_{\mathrm{cross}}(\theta,\phi).
\end{equation}
Defining $\bm{\gamma} =  [\gamma_1, ..., \gamma_{P}]^T$ and $\bm{\beta} = [\beta_1, ..., \beta_{P}]^T$ for $P$ arriving signals, we can construct the antenna response matrix
$
	\bm{A}(\bm{\theta},\bm{\phi},\bm{\gamma},\bm{\beta}) =
	\begin{bmatrix}
		\bm{a}(\theta_1,\phi_1,\gamma_1,\beta_1) & ... & \bm{a}(\theta_P,\phi_P,\gamma_P,\beta_P)
	\end{bmatrix}
$
and extend the signal model \cref{eq:multsignal} to
\begin{equation}\label{eq:polasignal}
	\bm{r}(n) = \bm{A}(\bm{\theta},\bm{\phi},\bm{\gamma},\bm{\beta}) \bm{s}(n) + \bm{w}(n).
\end{equation}
Similar to \cref{eq:cohml}, the polarimetric \ac{ML} estimator (P-ML) is given by
\begin{equation}\label{eq:polml}
	[\hat{\bm{\theta}},\, \hat{\bm{\phi}},\, \hat{\bm{\gamma}},\, \hat{\bm{\beta}}]	= \arg \min_{[\bm{\theta},\, \bm{\phi},\, \bm{\gamma},\, \bm{\beta}]} \operatorname{Re}\{\operatorname{tr}\{\bm{\Pi_{A}}^\perp \hat{\bm{R}_r}\}\}.
\end{equation}
It is assumed that the polarization parameters are stationary during the observation time. The \ac{CRB} matrix \cref{eq:cohcrb} is extended for the polarimetric \ac{CRB} (P-CRB) to
\begin{equation}\label{eq:polcrb}
	\text{CRB}[\hat{\bm{\theta}},\, \hat{\bm{\phi}},\, \hat{\bm{\gamma}},\, \hat{\bm{\beta}}] = \frac{\sigma^2}{2N} \operatorname{Re}\{\bm{D}^H \bm{\Pi_A}^\perp \bm{D} \odot \bm{Z}^T \bm{R}_{\bm{s}} \bm{Z}\}^{-1}.
\end{equation}
Setting $\bm{a}_p = \bm{a}(\theta_p,\phi_p,\gamma_p,\beta_p)$,
\begin{equation}
	\bm{D} = 
	\begin{bmatrix} ... &
		\frac{\partial \bm{a}_p}{\partial\theta_p} & ... &
		\frac{\partial \bm{a}_p}{\partial\phi_p} & ... &
		\frac{\partial \bm{a}_p}{\partial\gamma_p} & ... &
		\frac{\partial \bm{a}_p}{\partial\beta_p} & ... 
	\end{bmatrix}
\end{equation}
is defined analogously to \cref{eq:Dc} for $p=1,...,P$ and
$\bm{Z} = \begin{bmatrix}
	\mathbb{I}_P & \mathbb{I}_P & \mathbb{I}_P & \mathbb{I}_P
\end{bmatrix}
$.

The estimators introduced in this section, as well as their corresponding error bounds in terms of the \ac{CRB}, are summarized in \cref{tab.estimators}. Each of them requires the application of either the \ac{AIT}, see \cref{ss:ait}, or \ac{WM}, see \cref{ss:wms}.
\begin{table}[t]
	\caption{Estimators and corresponding \aclp{CRB}.}
	\centering
	\begin{tabular}{lll}
		\toprule
		\textbf{Signal model} & \textbf{Estimator} & \textbf{\ac{CRB}} \\ 
		\midrule
		Non-coherent & NC-ML \cref{eq:noncohml} & NC-CRB \cref{eq:noncohcrb,eq:noncohun} \\
		& NC-RC \cref{eq:noncohlc} & NC-RC-CRB \cref{eq:noncohcrb,eq:noncohlcun} \\
		\midrule
		Coherent & C-ML \cref{eq:cohml} & C-CRB \cref{eq:cohcrb} \\
		\midrule
		Polarimetric & P-ML \cref{eq:polml} & P-CRB \cref{eq:polcrb} \\
		\bottomrule
	\end{tabular}
	\label{tab.estimators}
\end{table}

\section{Performance Analysis}\label{s:performance}
\subsection{2D DoA Estimation}\label{ss:2d}
\begin{figure}[t]
	\centering
	\includegraphics{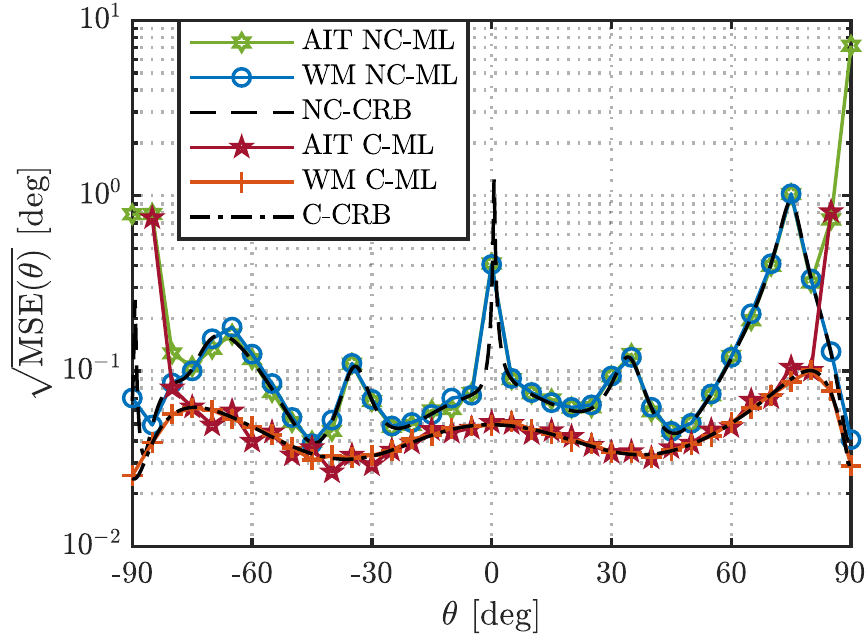}
	\caption{Simulated \ac{RMSE} of \ac{AIT} and \ac{WM} applied for non-coherent estimation \cref{eq:noncohml} and coherent estimation \cref{eq:cohml} and their respective \acp{CRB}, depending on the \ac{DoA} $\theta$, for $\text{SNR} = \unit[20]{dB}$.}
	\label{fig:MSE_theta_aitwm}
\end{figure}
Simulations have been performed to assess the \ac{DoA} estimation performance using a single \ac{MMA} with the aim to compare \ac{AIT} and \ac{WM} and the different estimation schemes. The investigated \ac{MMA} prototype has been presented in \cite{manteuffel2016} as part of an array of \acp{MMA}. For the simulations, the received signals are generated based on the signal models \cref{eq:signal,eq:multsignal} and the antenna response vector $\bm{a}(\theta,\phi)$ is given by the original \ac{EMF} simulation data with a $5^\circ$ grid. In this section we focus on 2D \ac{DoA} estimation for a transmitter located in the x-z-plane of the antenna, i.e. $\phi=0^\circ$.
The number of samples used is always $N=1000$ and we evaluate the \ac{DoA} estimation performance in terms of $\sqrt{\mathrm{MSE}(\theta)} = \sqrt{\frac{1}{N_{\mathrm{mc}}}\sum_{n_{\mathrm{mc}}=1}^{N_{\mathrm{mc}}} (\hat{\theta}_{n_{\mathrm{mc}}}-\theta)^2}$  for $N_{\mathrm{mc}}=1000$ Monte Carlo runs. For the non-coherent estimation scheme the relation between signal power $\check{s}$ and noise variance $\sigma^2$ is nonlinear. For the simulations, a fixed $\sigma^2 = k_{\mathrm{B}} T B$ with Boltzmann constant $k_{\mathrm{B}}$, noise temperature $T = \unit[290]{K}$ and bandwidth $B = \unit[1]{MHz}$ has been used.

First we want to analyze the \ac{DoA} estimation performance using \ac{AIT}, see \cref{ss:ait}, and \ac{WM}, see \cref{ss:wms}, as models for the \ac{MMA} response vector. \Cref{fig:MSE_theta_aitwm} shows the \acf{RMSE} for coherent and non-coherent estimation with both models depending on the \ac{DoA} $\theta$. Using \ac{AIT}, the \ac{RMSE} for non-coherent and coherent estimation is close to the \ac{CRB} for the main beam of the antenna. For lower elevations however, i.e. $\theta \geq 30^\circ$, \ac{AIT} suffers from additional errors due to a model mismatch between true and approximated \ac{MMA} response. Applying \ac{WM}, the \ac{RMSE} always approaches the \ac{CRB}. Therefore, when \ac{WM} is applied with a sufficient number of coefficients, it is able to perfectly interpolate the antenna response at the provided spatial sampling points.

\begin{figure}[t]
	\centering
	\includegraphics{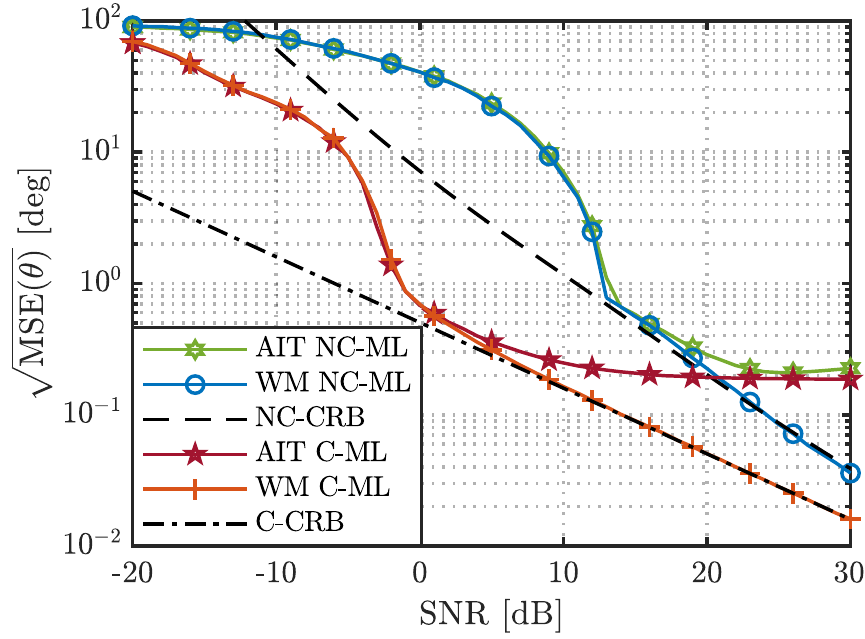}
	\caption{Simulated \ac{RMSE} of \ac{AIT} and \ac{WM} applied for non-coherent estimation \cref{eq:noncohml} and coherent estimation \cref{eq:cohml} and their respective \acp{CRB}, averaged over $\theta$.}
	\label{fig:MSE_SNR_aitwm}
\end{figure}
In \cref{fig:MSE_SNR_aitwm} we show the \ac{RMSE} for coherent and non-coherent estimation with \ac{AIT} and \ac{WM}, averaged over $\theta \in [-90^\circ,90^\circ]$. The received signals are again based on the original \ac{EMF} data. In the lower \ac{SNR} regime, \ac{AIT} and \ac{WM} show similar performance. Contrarily for high \ac{SNR}, \ac{WM} asymptotically approaches the respective \ac{CRB}, whereas using \ac{AIT}, the \ac{RMSE} does not drop below an error floor of $0.5^\circ$ for non-coherent and $0.3^\circ$ for coherent estimation. As stated already in the last paragraph, \ac{AIT} suffers from additional errors in the high \ac{SNR} domain due to a model mismatch between true and approximated \ac{MMA} response. From here on we present only results using \ac{WM}, as it achieves exact interpolation with the employed number of coefficients. Nevertheless \ac{AIT} is also a suitable approach for many applications, see the discussion in \cref{s:discussion}.

\begin{figure}[t]
	\centering
	\includegraphics{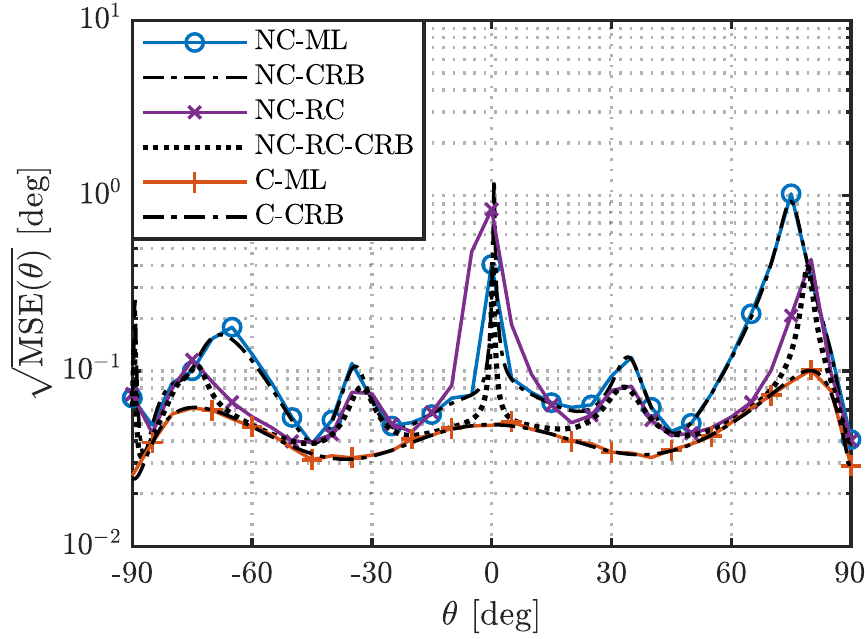}
	\caption{Simulated \ac{RMSE} of non-coherent \ac{ML} estimator \cref{eq:noncohml}, non-coherent \ac{RC} estimator \cref{eq:noncohlc}, coherent \ac{ML} estimator \cref{eq:cohml} and their respective \acp{CRB} depending on the \ac{DoA} $\theta$ for $\text{SNR} = \unit[20]{dB}$.}
	\label{fig:MSE_theta_coh_noncoh}
\end{figure}
\Cref{fig:MSE_theta_coh_noncoh} shows the \ac{RMSE} for 2D \ac{DoA} estimation with the non-coherent, i.e. \ac{RSS} based, and coherent estimators. For the given \ac{SNR} of $\unit[20]{dB}$, the non-coherent \ac{ML} estimator approaches the corresponding \ac{CRB}. The \ac{CRB} however, i.e. the achievable estimator performance, depends strongly on $\theta$. For the simulation of the \ac{RC} estimator, $\hat{\sigma}^2$ was estimated from $N=1000$ samples where no signal was present, corresponding to e.g. an unoccupied \ac{TDMA} slot. The \ac{RMSE} of the \ac{RC} is most of the time close to the corresponding \ac{CRB}, except around $\theta=0^\circ$. There the approximations leading from \cref{eq:noncohlh} to \cref{eq:noncohlcnonred} cause additional errors, likely due to large gain differences between the ports. The coherent \ac{ML} estimator approaches the \ac{CRB} for all $\theta$ values and the performance is relatively constant over $\theta$ compared to the non-coherent \ac{ML} estimator.

\begin{figure}[t]
	\centering
	\includegraphics{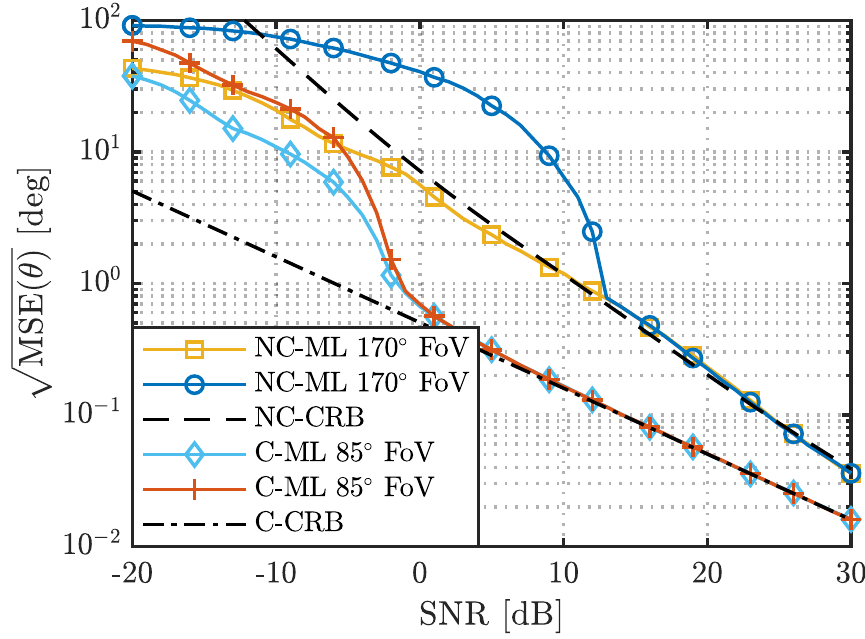}
	\caption{Simulated \ac{RMSE} of non-coherent \ac{ML} estimator \cref{eq:noncohml}, coherent \ac{ML} estimator \cref{eq:cohml} and their respective \acp{CRB}. For $170^\circ$ \acf{FoV} the \ac{RMSE} is calculated over $\theta \in [-85^\circ,85^\circ]$, for $85^\circ$ \ac{FoV} over $\theta \in [-85^\circ,0^\circ]$ and $\theta \in [0^\circ,85^\circ]$.}
	\label{fig:MSE_SNR_coh_noncoh}
\end{figure}
\begin{figure}[t]
	\centering
	\includegraphics{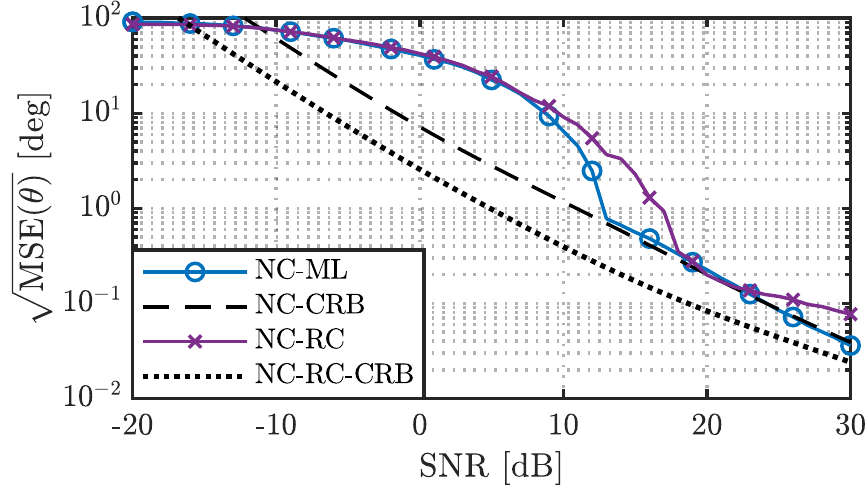}
	\caption{Simulated \ac{RMSE} of non-coherent \ac{ML} estimator \cref{eq:noncohml}, non-coherent \ac{RC} estimator \cref{eq:noncohlc} and their respective \acp{CRB}, averaged over $\theta$.}
	\label{fig:MSE_SNR_noncoh_lc}
\end{figure}
In \cref{fig:MSE_SNR_coh_noncoh} the \ac{RMSE} is calculated over $\theta \in [-85^\circ,85^\circ]$ for $170^\circ$ \acf{FoV}, and over $\theta \in [-85^\circ,0^\circ]$ and $\theta \in [0^\circ,85^\circ]$ for $85^\circ$ \ac{FoV} and plotted versus \ac{SNR}.
According to \cref{fig:Power_2D_RHCP}, the \ac{MMA} power pattern is relatively symmetric with respect to $\theta=0^\circ$, thus the discrimination between positive and negative values of $\theta$ is handicapped. \Cref{fig:MSE_SNR_coh_noncoh} shows that the non-coherent estimator for $170^\circ$ \ac{FoV} has much higher errors in the low \ac{SNR} regime compared to $85^\circ$ \ac{FoV}. Above approximately \unit[13]{dB} \ac{SNR} both curves asymptotically approach the \ac{CRB}. The coherent estimator asymptotically approaches its \ac{CRB} above $\text{SNR} = \unit[7]{dB}$. For lower \ac{SNR}, the difference between $170^\circ$ \ac{FoV} and $85^\circ$ \ac{FoV} is much smaller compared to the non-coherent case. The coherent estimator can efficiently use the phase response of the \ac{MMA}, see \cref{fig:Phase_2D_RHCP}, to distinguish between positive and negative $\theta$ and thus suffers less from estimation ambiguities.

\Cref{fig:MSE_SNR_noncoh_lc} shows the \ac{RMSE} averaged over $\theta \in [-85^\circ, 85^\circ]$ versus the \ac{SNR} for the non-coherent \ac{ML} and \ac{RC} estimator. For the \ac{RC} estimator, $\hat{\sigma}^2$ was again estimated from $N=1000$ noise samples. The performance of \ac{ML} and \ac{RC} estimator is similar, however \ac{ML} asymptotically approaches its \ac{CRB}, whereas \ac{RC} does not. Given its lower complexity, \ac{RC} presents a viable alternative to \ac{ML} for non-coherent estimation.

\begin{figure}[t]
	\centering
	\includegraphics{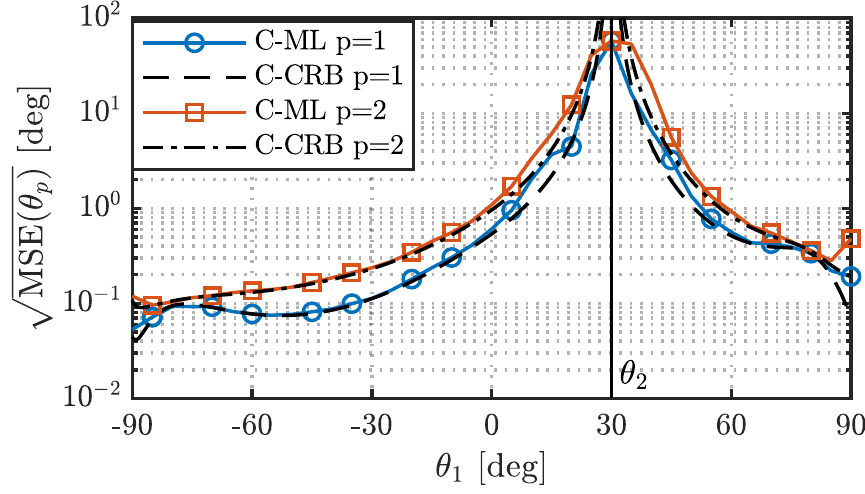}
	\caption{Simulated \ac{RMSE} of coherent \ac{ML} estimator \cref{eq:cohml} and \ac{CRB} for two signals arriving from directions $\theta_1$ and $\theta_2$. The second signal has \unit[6]{dB} less power.}
	\label{fig:MSE_theta_cohP2}
\end{figure}
In \cref{fig:MSE_theta_cohP2} we examine the case of $P=2$ incoming signals for the coherent ML estimator \cref{eq:cohml}. The first signal arrives from a variable angle $\theta_1 \in [-90^\circ, 90^\circ]$, while the second signal arrives from $\theta_2 = 30^\circ$ with \unit[6]{dB} less power. This case where $P>1$ is common in practice due to multipath propagation of radio signals. Often there is not only the line-of-sight signal arriving at the receiver, but also several multipath signals from reflection and scattering. The multipath signals are delayed and usually attenuated with respect to the line-of-sight signal. The plot reveals that the two signals can be separated well and the \ac{RMSE} of the estimator is close to the respective \ac{CRB}, unless the two signals are very close together. When the two signals get closer, they become more correlated and separation becomes more challenging, which can be seen in the increasing \ac{RMSE} and \ac{CRB}. In the limit, for very close spacing, separation is not possible any more.

\subsection{3D DoA Estimation}\label{ss:3d}
\begin{figure}[t]
	\centering
	\subfloat[$\sqrt{\mathrm{CRB}(\theta)}$ (left) and $\sqrt{\mathrm{CRB}(\phi)}$ (right)]{\includegraphics{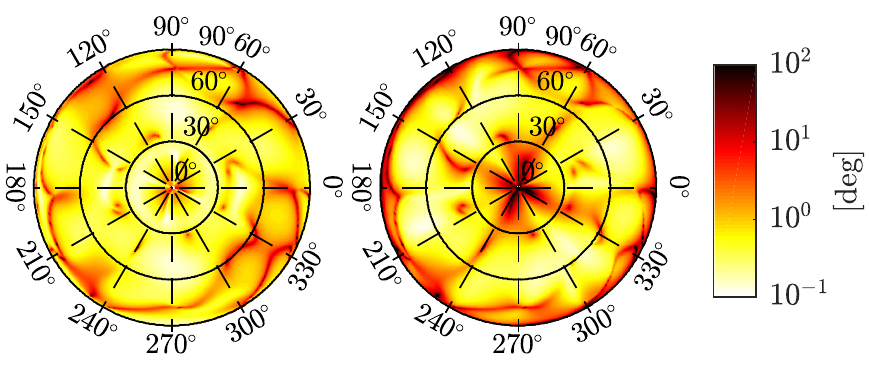}%
		\label{fig:surfNoncohCRB}}
	\hfil
	\subfloat[$\sqrt{\mathrm{MSE}(\theta)}$ (left) and $\sqrt{\mathrm{MSE}(\phi)}$ (right)]{\includegraphics{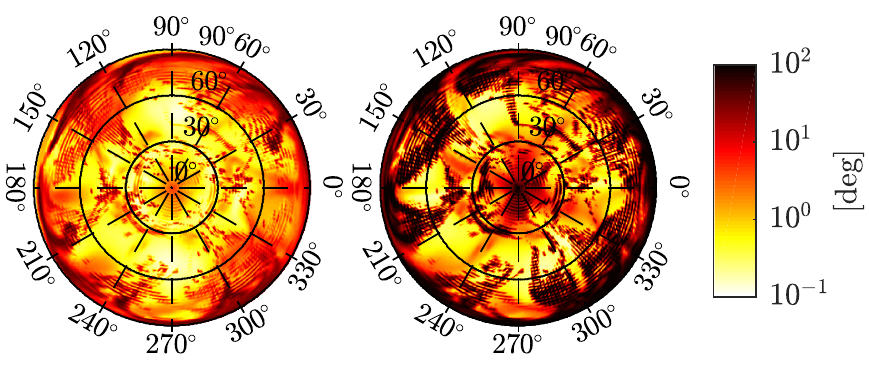}%
		\label{fig:surfNoncohMSE}}
	\hfil
	\subfloat[$\sqrt{\mathrm{MSE}(\theta)} / \sqrt{\mathrm{CRB}(\theta)}$ (left) and $\sqrt{\mathrm{MSE}(\phi)} / \sqrt{\mathrm{CRB}(\phi)}$ (right)]{\includegraphics{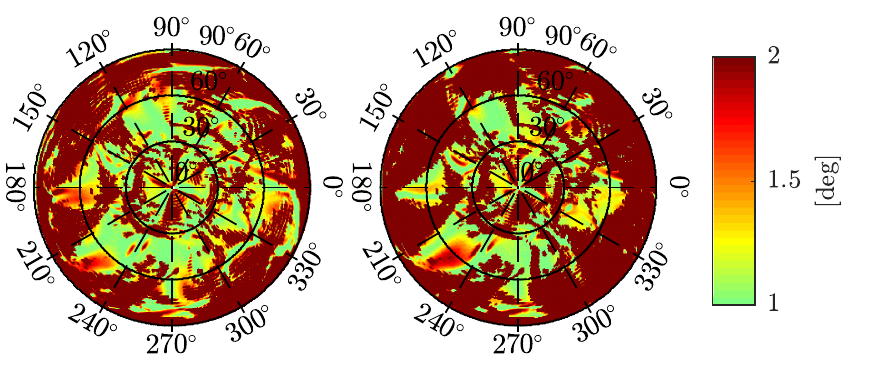}%
		\label{fig:surfNoncohRatio}}
	\caption{Simulated \ac{RMSE} for the non-coherent \ac{ML} estimator \cref{eq:noncohml} and corresponding \ac{CRB} \cref{eq:noncohcrb,eq:noncohun} with $\text{SNR} = \unit[10]{dB}$. Radius represents $\theta \in [0^\circ,\, 90^\circ]$, angle represents $\phi \in [0^\circ,\, 360^\circ)$.}
	\label{fig:surfNoncoh}
\end{figure}
For the 3D section we use \ac{WM} with spherical harmonic functions \cref{eq:sh} as basis and $U = 144$ coefficients to generate the received signals based on \cref{eq:signal,eq:multsignal}, which allows to show results for a finer $\theta$ and $\phi$ grid of $1^\circ$ compared to the original \ac{EMF} simulation data with $5^\circ$ grid. For the estimator we use spherical harmonic functions with $U = 64 \approx 8 \kappa^2 R_{\mathrm{s}}^2 + 4 \kappa R_{\mathrm{s}} +1$  coefficients.

In \cref{fig:surfNoncoh} the simulated RMSE and the corresponding \ac{CRB} for $\theta$ and $\phi$, i.e. 3D, non-coherent \ac{DoA} estimation are shown. Similar to \cref{fig:MSE_theta_coh_noncoh} for the 2D case, the \ac{CRB} shown in \cref{fig:surfNoncohCRB} for 3D varies depending on $\theta$ and $\phi$. \cref{fig:surfNoncohRatio} shows the ratio between the simulated RMSE of the non-coherent \ac{ML} estimator in \cref{fig:surfNoncohMSE} and its corresponding \ac{CRB} in \cref{fig:surfNoncohCRB}. The estimator approaches the \ac{CRB} when the ratio is close to one. For the non-coherent estimator this is only the case for some angles with $\theta < 60^\circ$. There are also many angles with $\theta < 60^\circ$ where the \ac{CRB} is not approached. For $\theta \geq 60^\circ$, excessive estimation errors occur. One explanation is the antenna power pattern, see \cref{fig:Power_3D_RHCP,fig:Power_2D_RHCP}. For $\theta$ approaching $90^\circ$, the antenna gain is very low, leading to a low \ac{SNR}. Another explanation, which is also discussed in \cref{ss:2d} and later in this section, is that the non-coherent estimator suffers from estimation ambiguities, which apparently are more harmful in 3D than in 2D estimation.

\begin{figure}[t]
	\centering
	\subfloat[$\sqrt{\mathrm{CRB}(\theta)}$ (left) and $\sqrt{\mathrm{CRB}(\phi)}$ (right)]{\includegraphics{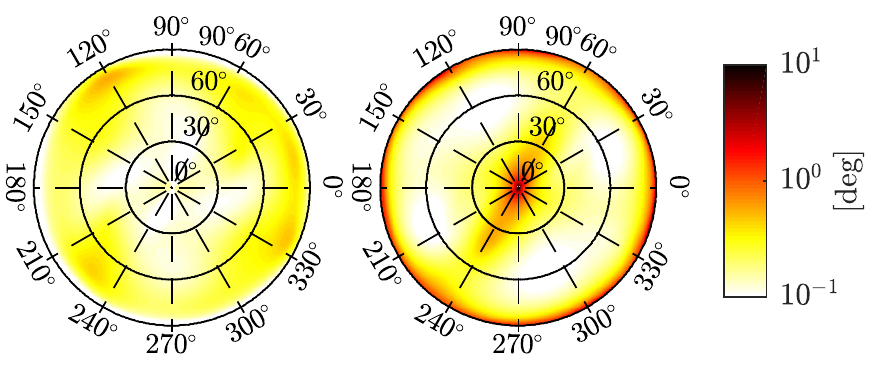}%
		\label{fig:surfCohCRB}}
	\hfil
	\subfloat[$\sqrt{\mathrm{MSE}(\theta)}$ (left) and $\sqrt{\mathrm{MSE}(\phi)}$ (right)]{\includegraphics{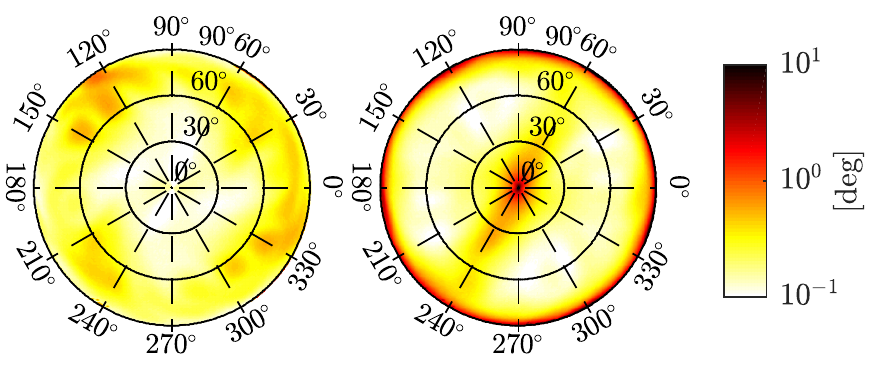}%
		\label{fig:surfCohMSE}}
	\hfil
	\subfloat[$\sqrt{\mathrm{MSE}(\theta)} / \sqrt{\mathrm{CRB}(\theta)}$ (left) and $\sqrt{\mathrm{MSE}(\phi)} / \sqrt{\mathrm{CRB}(\phi)}$ (right)]{\includegraphics{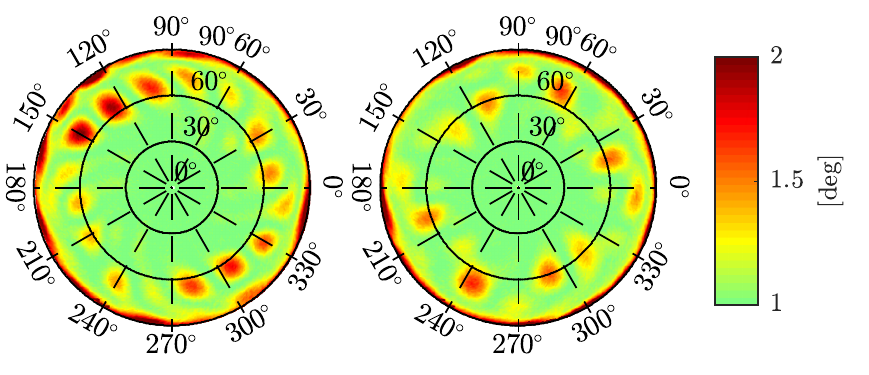}%
		\label{fig:surfCohRatio}}
	\caption{Simulated \ac{RMSE} for the coherent \ac{ML} estimator \cref{eq:cohml} and corresponding \ac{CRB} \cref{eq:cohcrb} with $\text{SNR} = \unit[10]{dB}$. Radius represents $\theta \in [0^\circ,\, 90^\circ]$, angle represents $\phi \in [0^\circ,\, 360^\circ)$.}
	\label{fig:surfCoh}
\end{figure}
\Cref{fig:surfCoh} shows RMSE and \ac{CRB} for 3D coherent \ac{DoA} estimation. As expected, the \ac{CRB} in \cref{fig:surfCohCRB} is lower and more uniform compared to the non-coherent case in \cref{fig:surfNoncohCRB}. \Cref{fig:surfCohRatio}, showing the ratio between the simulated RMSE of the coherent estimator, see \cref{fig:surfCohMSE}, and its corresponding \ac{CRB}, see \cref{fig:surfCohCRB}, reveals that the RMSE of the coherent estimator approaches the \ac{CRB} for $\theta < 45^\circ$. For $\theta \geq 45^\circ$, some angles with ratio greater than one are visible. Unlike for the non-coherent estimator, the deviation from the \ac{CRB} is small. The coherent estimator outperforms the non-coherent one also for 3D, and is able to provide \ac{DoA} estimates with sub-degree accuracy, except for very low elevations.

\begin{figure}[t]
	\centering
	\includegraphics{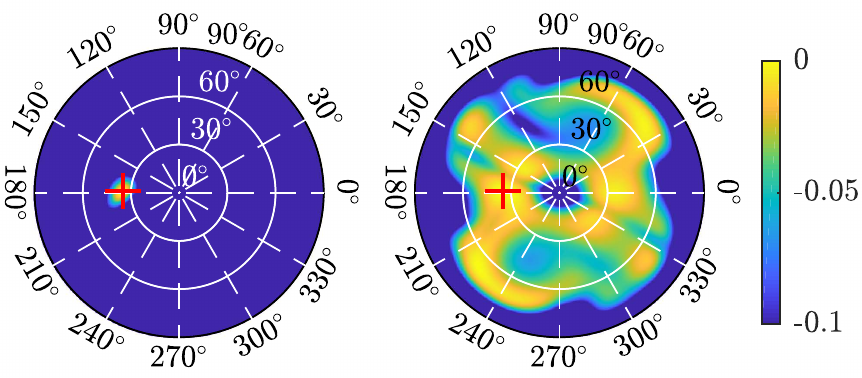}
	\caption{Normalized $[-1, 0]$ log-likelihood functions for the C-ML \cref{eq:cohlh}, left, and NC-ML estimator \cref{eq:noncohlh}, right. The signal is coming from $\theta = 35^\circ$, $\phi = 178^\circ$ (red cross) with $\text{SNR} = \unit[15]{dB}$.  Radius represents $\theta \in [0^\circ,\, 90^\circ]$, angle represents $\phi \in [0^\circ,\, 360^\circ)$.}
	\label{fig:logLikelihood}
\end{figure}
\Cref{fig:logLikelihood} shows the log-likelihood functions for the coherent \cref{eq:cohlh}, and non-coherent estimator \cref{eq:noncohlh} for a fixed \ac{DoA}. This plot reveals where the significant difference between non-coherent and coherent estimator comes from. The coherent log-likelihood function has only one sharp peak at the true \ac{DoA}. In contrast to that, the non-coherent log-likelihood function has multiple peaks. In the presence of noise, the estimator may lock to the wrong peak being a local maximum. That causes a deviation of the non-coherent estimator from the \ac{CRB}, see \cref{fig:surfNoncoh}.

\begin{figure}[t]
	\centering
	\includegraphics{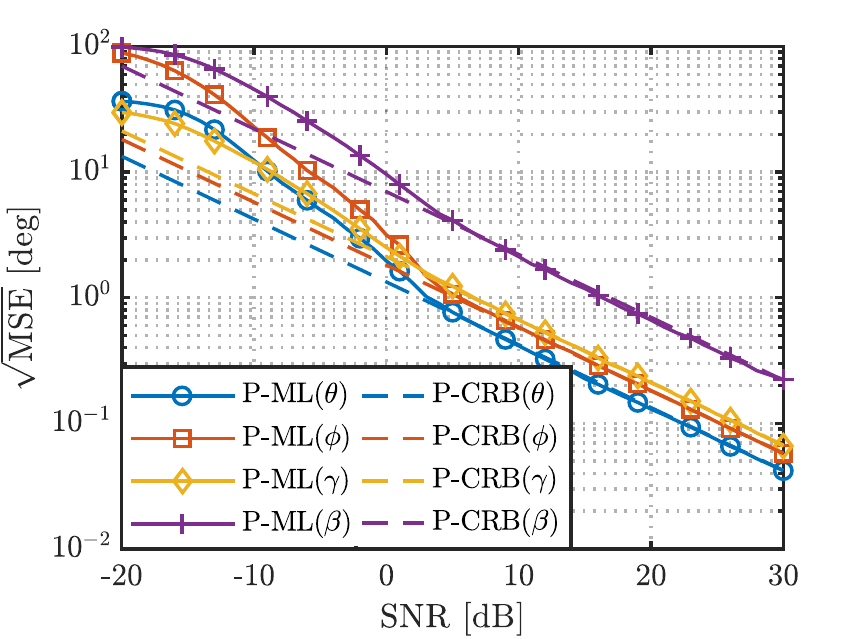}
	\caption{Simulated \ac{RMSE} of polarimetric \ac{ML} estimator \cref{eq:polml} with corresponding \acp{CRB}, averaged over $\theta$, $\phi$, $\gamma$, $\beta$.}
	\label{fig:MSE_SNR_pol}
\end{figure}
Finally we want to assess the performance of the joint \ac{DoA} and polarization estimation approach introduced in \cref{ss:pol}. The RMSE of the polarimetric \ac{ML} estimator versus the \ac{SNR} is visualized in \cref{fig:MSE_SNR_pol}. The results are averaged over the \ac{DoA} parameters $\theta \in [0^\circ,80^\circ)$, $\phi \in [0^\circ,360^\circ)$ and the polarization parameters $\gamma \in [10^\circ,80^\circ]$ and $\beta \in [-180^\circ,180^\circ)$. Above an \ac{SNR} of \unit[2]{dB}, the RMSE of all parameter estimates asymptotically approach the \ac{CRB}. Therefore, using the investigated \ac{MMA} prototype a determination of the signal polarization is deemed possible.

\section{Discussion}\label{s:discussion}
Two different approaches for signal processing with \acp{MMA} are presented in this paper, one based on \ac{AIT}, see \cref{ss:ait}, and one based on \ac{WM}, see \cref{ss:wms}. \ac{AIT} clearly has an advantage when real-time processing is required. It allows to transform the received signals into the domain of an ideal array with a certain geometry, e.g. uniform linear or rectangular. Efficient algorithms like \ac{ESPRIT} \cite{roy1986} or unitary \ac{ESPRIT} \cite{haardt1995} can be applied. A challenge for the method is when multiple signals arrive at different sectors, as this leads to out-of-sector errors and degraded estimation performance. \ac{WM} does not suffer from this drawback, because it does not require sectorization. However its computational cost is in general higher, since low complexity algorithms limited to uniform linear/rectangular array geometries cannot be used. In \cite{costa2012} a \ac{MUSIC} variant based on \ac{WM} is presented as a method with moderate complexity, but it faces difficulties in the case of coherent signals. 

The \ac{WM} technique shown in \cref{ss:wms} allows to use different basis functions for the expansion. For 3D, spherical harmonic functions \cref{eq:Ylm} or Fourier functions \cref{eq:eadf} can be used. Both variants are equivalently valid and it has even been shown that one can be transformed into the other \cite{costa2010}. For spherical harmonics, less coefficients are necessary for an accurate interpolation, but the computational complexity is higher because the evaluation of the associated Legendre polynomial $P_l^m(.)$ in \cref{eq:Ylm} is costly. On the other hand, Fourier functions require more coefficients, but they can be efficiently evaluated with the FFT. Care has to be taken because they are orthonormal on the torus, not on the sphere \cite{costa2010}. This has to be taken into account when performing the expansion.

\section{Conclusion}\label{s:conclusion}
This paper addresses the question how \acfp{MMA} can be used for \ac{DoA} estimation. We define an \ac{MMA} as a multiport antenna, where different characteristic modes are excited independently. \acp{MMA} have so far been designed and investigated only for communications, while their potential for positioning has not been leveraged. To enable \ac{DoA} estimation with \acp{MMA}, we present two suitable ways, based on either \acf{AIT} or \acf{WM}. Both fully take antenna nonidealities like mutual coupling into account. We further show how non-coherent, i.e. \ac{RSS} based, coherent and joint \ac{DoA} and polarization estimation can be carried out. Based on \ac{EMF} simulation data, we perform extensive simulations in both 2D and 3D to assess the expected performance. We compare \ac{AIT} and \ac{WM} in terms of \ac{DoA} estimation performance and show that \ac{WM} has an advantage in the high \ac{SNR} regime. For low-cost and low-complexity receivers, non-coherent \ac{DoA} estimation based on \ac{RSS} measurements is possible. However it suffers from estimation ambiguities, especially in the 3D case, and thus requires a relatively high \ac{SNR} for accurate results. The standard coherent approach does not suffer from this problem and performs better. The coherent receiver achieves sub-degree accuracy for a 2D scenario with an \ac{SNR} above \unit[5]{dB}, whereas the non-coherent one requires at least \unit[14]{dB}.  As the investigated \ac{MMA} prototype features diverse polarizations, we also show that the polarization parameters of the incoming wave can be estimated. In conclusion, \ac{DoA} estimation with \acp{MMA} is both feasible and accurate. \acp{MMA} thus offer an appealing alternative to conventional antenna arrays, especially in applications with tight shape constraints.

\appendices
\crefalias{section}{appsec}
\section{Legendre Polynomials and Derivatives of Complex Spherical Harmonics}\label{a:sh}
The spherical harmonics \cref{eq:Ylm} with degree $l$ and order $m$ can be calculated with the associated Legendre polynomial \cite{olver2010}
\begin{equation}\label{eq:Plm}
P_l^m(x) = (-1)^m (1-x^2)^{m/2} \frac{\mathrm{d}^m}{\mathrm{d} x^m} P_l(x)
\end{equation}
and the Legendre polynomial
\begin{equation}
P_l(x) = \frac{1}{2^l \, l!} \frac{\mathrm{d}^l}{\mathrm{d} x^l} (x^2-1)^l.
\end{equation}
The partial derivatives of the spherical harmonics \cref{eq:Ylm} with respect to $\theta$ and $\phi$ are
\begin{subequations}
	\begin{align}
	&\frac{\partial}{\partial\theta} Y_l^m(\theta,\phi) = m\cot(\theta) Y_l^m(\theta,\phi) + \\
	&\qquad \sqrt{(l-m)(l+m+1)} e^{-j\phi} Y_l^{m+1}(\theta,\phi), \nonumber \\
	&\frac{\partial}{\partial\phi} Y_l^m(\theta,\phi) = jm Y_l^m(\theta,\phi).
	\end{align}
\end{subequations}

\section{Real Spherical Harmonics and Their Derivatives}\label{a:rsh}
The real version of the spherical harmonic functions, which can be applied in \cref{eq:sh} for the non-coherent signal model described in \cref{ss:noncoh}, are given by
\begin{equation}\label{eq:realsh}
	Y_l^m(\theta,\phi) = 
	\begin{cases}
		\sqrt{2} N_l^m \cos(m\phi) P_l^m(\cos(\theta)) & m>0 \\
		N_l^0 P_l^m(\cos(\theta)) & m=1 \\
		\sqrt{2} N_l^{|m|} \sin(|m|\phi) P_l^{|m|}(\cos(\theta)) & m<0, \\
	\end{cases}
\end{equation}
with degree $l = 0,...,L$, order $m = -l,...,l$ and $P_l^m(.)$ given by \cref{eq:Plm}. The normalization factor $N_l^m$ is defined as
\begin{equation}
	N_l^m = \sqrt{\frac{2l+1}{4\pi}\frac{(l-m)!}{(l+m)!}}.
\end{equation}
The derivative of the real spherical harmonics with respect to $\theta$ is given by
\begin{equation}
	\frac{\partial}{\partial\theta} Y_l^m(\theta,\phi) =
	\begin{cases}
		\sqrt{2} N_l^m \cos(m\phi) \frac{\partial P_l^m(\cos(\theta))}{\partial\theta} & m>0 \\
		N_l^0 \frac{\partial P_l^m(\cos(\theta))}{\partial\theta}  & m=1 \\
		\sqrt{2} N_l^{|m|} \sin(|m|\phi) \frac{\partial P_l^{|m|}(\cos(\theta))}{\partial\theta} & m<0. \\
	\end{cases}
\end{equation}
It contains the derivative of the associated Legendre polynomial \cite{olver2010}
\begin{equation}
\begin{split}
	&\frac{\partial P_l^m(\cos(\theta))}{\partial\theta} =\\
	&\quad 1+l-m \, \sin(\theta) P_{l+1}^m(cos(\theta)) - \frac{l+1}{\tan(\theta)}P_l^m(\cos(\theta)).
\end{split}
\end{equation}
The derivative of the real spherical harmonics with respect to $\phi$ is given by
\begin{equation}
\begin{split}
	&\frac{\partial}{\partial\phi} Y_l^m(\theta,\phi) =\\
	&\quad
	\begin{cases}
		\sqrt{2} N_l^m (-m)\sin(m\phi) P_l^m(\cos(\theta)) & m>0 \\
		0 & m=1 \\
		\sqrt{2} N_l^{|m|} (-m)\cos(m\phi) P_l^{|m|}(\cos(\theta)) & m<0. \\
	\end{cases}
\end{split}
\end{equation}

\section{Proof that \ac{RSS} Measurements $\check{\bm{r}}$ are Approximately Gaussian Distributed}\label{a:sm}
Here we show that the \ac{RSS} measurements $\check{\bm{r}} = [\check{r}_1,...,\check{r}_M]^T$, with $\check{r}_m$ given by \cref{eq:Prm}, can be approximated by a Gaussian distribution with mean \cref{eq:gmu} and covariance matrix \cref{eq:gsigma}.
For clarity we use scalar notation, the subscript $m$ refers to the $m$-th element of the respective vector. 
Defining $r_{m,\text{r}}(n) = \operatorname{Re}\{r_m(n)\}$ and $r_{m,\text{i}}(n) = \operatorname{Im}\{r_m(n)\}$, the sum of the squared magnitude of the received signal,
\begin{equation}
	\tilde{r}_m = \sum_{n=1}^{N} |r_m(n)|^2 = \sum_{n=1}^{N} r_{m,\text{r}}^2(n) + r_{m,\text{i}}^2(n) \sim \chi^2(2N,\Lambda,\sigma^2/2)
\end{equation}
follows a noncentral $\chi^2$ distribution \cite{proakis2008} with $2N$ degrees of freedom. The noncentrality parameter can be derived as
\begin{equation}\label{eq:lambda}
\begin{split}
	\Lambda &= \sum_{n=1}^{N} \left( \operatorname{E}\{r_{m,\text{r}}(n)\}^2 + \operatorname{E}\{r_{m,\text{i}}(n)\}^2 \right) \\
	&= \sum_{n=1}^{N}  \left(\operatorname{Re}\{a_m(\theta,\phi)s(n)\}^2 + \operatorname{Im}\{a_m(\theta,\phi)s(n)\}^2\right) \\
	&= \sum_{n=1}^{N} |a_m(\theta,\phi)|^2 |s(n)|^2 \\
	&= \sum_{n=1}^{N} g_m(\theta,\phi)  |s(n)|^2 \\
	&= N g_m(\theta,\phi) \check{s}.
\end{split}
\end{equation}
The \ac{PDF} of the noncentral $\chi^2$ distribution is given by
\begin{equation}
	p_{\tilde{r}_m}(x) = \frac{1}{\sigma^2} \left(\frac{x}{\Lambda}\right)^{\frac{N}{2}} e^{-\frac{\Lambda+x}{\sigma^2}} I_{N}\left(\frac{2\sqrt{\Lambda x}}{\sigma^2}\right),
\end{equation}
where $I_\nu(.)$ is the modified Bessel function of the first kind, see \cite{olver2010}. Since $\check{r}_m$ is just a scaled version of that, its distribution can be obtained by transformation $p_{\check{r}_m}(x)= N p_{\tilde{r}_m}(Nx)$.
By inserting \cref{eq:lambda}, we obtain the \ac{PDF}
\begin{equation}\label{eq:pdf}
\begin{split}
	p_{\check{r}_m}(x) = &\frac{N}{\sigma^2} \left(\frac{x}{g_m(\theta,\phi)\check{s}}\right)^{\frac{N}{2}} e^{-\frac{N(g_m(\theta,\phi)\check{s}+x)}{\sigma^2}}\\
	&I_{N}\left(\frac{2N\sqrt{g_m(\theta,\phi) \check{s}  x}}{\sigma^2}\right).
\end{split}
\end{equation}
The mean and variance can be derived as
\begin{equation}\label{eq:mu}
\begin{split}
	\tilde{\mu}_m
	&= \operatorname{E}[\check{r}_m] = N^{-1} \, \operatorname{E}\{\tilde{r}_m\} \\
	&= N^{-1} (N\sigma^2+\Lambda) \\
	&= g_m(\theta,\phi) \check{s} + \sigma^2,
\end{split}
\end{equation}
\begin{equation}\label{eq:sigma}
\begin{split}
	\tilde{\sigma}_m^2 &= \operatorname{var}\{\check{r}_m\} = N^{-2} \, \operatorname{var}\{\tilde{r}_m\} \\
	&= N^{-2} (N\sigma^4+2\sigma^2\Lambda) \\
	&= N^{-1}(\sigma^4+2g_m(\theta,\phi)\check{s}\sigma^2).
\end{split}
\end{equation}
For a growing number of samples $N$, \cref{eq:pdf} approaches a Gaussian distribution $\check{r}_m \sim \mathcal{N}(\tilde{\mu}_m, \tilde{\sigma}_m^2)$ due to the central limit theorem. The approximation is reasonable for $N>25$ \cite{box2005}.

\section*{Acknowledgment}
The authors would like to thank Dirk Manteuffel and his team for providing the antenna pattern of the \ac{MMA} prototype investigated in this paper. Fruitful discussions with Kazeem A. Yinusa are highly appreciated.

\ifCLASSOPTIONcaptionsoff
  \newpage
\fi



%
%
%

\bibliography{bibliography_abrv}
\bibliographystyle{IEEEtran}

%







\end{document}